\documentclass[11pt,a4paper]{article}
\usepackage{jcappub}

\definecolor{blue}{rgb}{0,0,1}
\definecolor{green}{rgb}{0,1,0}
\definecolor{turkos}{rgb}{0,0.4,0.4}
\definecolor{red}{rgb}{1,0,0}
\definecolor{darkblue}{rgb}{0.4,0.2,1}

\begin{document}

\title{Gravitational, shear and matter waves in Kantowski-Sachs cosmologies}
\author[a,b]{Zolt\'{a}n Keresztes,}
\author[c]{Mats Forsberg,}
\author[c]{Michael Bradley,}
\author[d,e,f]{Peter K.S. Dunsby,}
\author[a,b]{L\'{a}szl\'{o} \'{A}. Gergely,}

\affiliation[a]{Department of Theoretical Physics, University of Szeged, Tisza Lajos
krt 84-86, Szeged 6720, Hungary}
\affiliation[b]{Department of Experimental Physics, University of Szeged, D\'{o}m T%
\'{e}r 9, Szeged 6720, Hungary}
\affiliation[c]{Department of Physics, Ume\aa\ University, Sweden}
\affiliation[d]{Astrophysics, Cosmology and Gravity Centre (ACGC), University of Cape Town, Rondebosch 7701, Cape Town, South Africa}
\affiliation[e]{Department of Mathematics and Applied Mathematics, University of Cape
Town, South Africa, Rondebosch 7701, Cape Town, South Africa}
\affiliation[f]{South African Astronomical Observatory,  Observatory 7925, Cape Town, South Africa}
\emailAdd{zkeresztes@titan.physx.u-szeged.hu}
\emailAdd{forsberg.mats.a.b@gmail.com}
\emailAdd{michael.bradley@physics.umu.se}
\emailAdd{peter.dunsby@uct.ac.za}
\emailAdd{gergely@physx.u-szeged.hu}

\abstract{
A general treatment of vorticity-free, perfect fluid perturbations of Kantowski-Sachs models with a positive 
cosmological constant are considered within the framework of the 1+1+2 covariant decomposition of spacetime. 
The dynamics is encompassed in six evolution equations for six harmonic coefficients, 
describing gravito-magnetic, kinematic and matter perturbations, while a set of algebraic 
expressions determine the rest of the variables. The six equations further decouple into a set of four 
equations sourced by the perfect fluid, representing forced oscillations and two uncoupled damped oscillator 
equations. 
 The two gravitational degrees of freedom are represented by pairs of gravito-magnetic perturbations. 
In contrast with the Friedmann case one of them is coupled to the matter density perturbations, becoming decoupled 
only in the geometrical optics limit. In this 
approximation, the even and odd tensorial perturbations of the Weyl tensor 
evolve as gravitational waves on the anisotropic Kantowski-Sachs background, while the modes describing the 
shear and the matter density gradient are out of phase dephased by $\pi /2$ and share the same speed of sound. 
}

\arxivnumber{1507.08300}
\maketitle
\section{Introduction}

The observed large scale distribution of galaxies, the fluctuations about
the isotropic cosmic microwave background radiation and the late time
acceleration of the universe seems to be well described by the $\Lambda $CDM
model, which is based on the assumption that the geometry of the universe is
given by the Robertson-Walker metric - see e.g., \cite{Komatsu49, Spergel48,
WMAP9yr, Planck1, Planck2}. However this fit is not perfect \cite{Bennett46,
Oliveria45, Vielva59, PlanckAnomaly} and because more than 95\% of the
matter budget needs to be described by the dark sector, it is worth
exploring what effect alternative cosmological models have on the basic
properties of the Universe \cite{alt1, alt2, alt3, alt4, alt5, alt6, alt7,
alt8, alt9, alt10, alt11, alt12}.

It is known that anisotropies in the Hubble and deceleration parameters
cannot be excluded by present observations \cite{H1}, \cite{H2} and \cite%
{Dec}. Consequently, perturbations of
anisotropic cosmological models have been considered by many authors, e.g., 
\cite{Doroschkevich, Perko, Hu, Abbott, Tomita, Gumruk, Pereira, Pitrou},
using both gauge dependent methods (e.g., \cite{Lifshitz}) or Bardeen's
gauge invariant formalism \cite{Bardeen}. For example the
perturbations of homogeneous and anisotropic universe of the Bianchi I type
was investigated in \cite{Pereira, Pitrou} by using Bardeen's
gauge-invariant method. However, the variables in Bardeen's theory are
defined with respect to a particular coordinate system, making their
geometrical and physical meaning not very transparent See the
discussion in \cite{Stewart}. By using a covariant approach, one
circumvents these problems by using the spatial curvature rather than the
metric as the defining variables \cite{Hawking, Olson}. In this way, a set
of gauge-invariant perturbation variables can be easily identified as the
ones that vanish on the chosen background \cite{StewartWalker, cov1, cov2,
cov3, cov4, cov5, cov6, cov7}. This feature of the covariant
approach makes it a very versatile method for studying perturbations on a
variety of backgrounds and physical situations and relating the results
obtained in a unified way \cite{cov-applications1,cov-applications2,cov-applications3,cov-applications4,cov-applications5}

In this paper we present for the first time a general treatment of the
vorticity-free perturbations of Kantowski-Sachs cosmologies with positive
cosmological constant, extending earlier work \cite{KASAscalar}, which
focused only on the scalar perturbation sector. Here we present
for the first time an analysis of a full scalar, vectorial and tensorial
perturbations, focusing on gravitational and matter wave evolutions.

In order to achieve this, we use a covariant and gauge invariant method \cite%
{Cargese}, in which spacetime is first split into a 1+3 form. The formalism {%
has been mainly employed for computation of cosmological
perturbations on a Friedmann background, applying the standard decomposition
theorems \cite{Stewart} (see for example Refs. \cite{cov1, cov3, perturb2,
perturb4}). If the 3-space at each point has a unique preferred direction, a
further decomposition of the spacetime into a 1+(1+2) form is useful in
situations where spacetime admits a spherical or Locally
Rotational (LRS) symmetry. This was first employed in Ref. \cite{LRS}, where
the spatial direction was singled out by local rotational symmetry (LRS).
The formalism was developed with the purpose of investigating general
gauge-invariant perturbations of the vacuum Schwarzschild spacetime \cite
{Schperturb}. With the further generalisation presented in Refs. \cite
{LRSIIscalar} and \cite{LRSIItensor}, it became possible to
describe gauge-invariant perturbations of LRS class II spacetimes for which
the complete set of evolution and constraint equations are given in Ref. 
\cite{1+1+2}.

In this paper, the variables describing an almost
Kantowski-Sachs spacetime are expanded into harmonics. We find that the
perturbation dynamics is described by six evolution equations for six
harmonic coefficients, together with a set of algebraic expressions, which
determine the evolution of the rest of the variables. The evolution
equations can be split into two sets - four which are sourced by the perfect
fluid, representing forced oscillations and the remaining two describing
damped oscillating gravito-magnetic perturbations. We further analyse the
equations using the geometrical optics approximation and find that four of
the gravito-magnetic quantities evolve as gravitational waves propagating on
the anisotropic Kantowski-Sachs background, while the shear and the matter
density gradients, which are out of phase by $\pi /2$, share
the same speed of sound.

The paper is organised as follows. In Section \ref%
{decomp_sec} we briefly review the 1+3 and 1+1+2 covariant approaches. In
Section \ref{KSbackqround_sec} a Kantowski-Sachs type background filled with
a perfect fluid is introduced. The equations governing the linear
perturbations are derived in Section \ref{perturb_sec}. All type of
perturbations (scalar, vector and tensor) are investigated, 
however they are restricted by vanishing anisotropic pressure and energy flux,
which mean that we assume a perfect fluid and that the 1+3 split is done with respect to the 4-velocity also in the perturbed spacetime. 
For simplicity we also choose to put the vorticity to zero. This imply
that the hypersurfaces perpendicular to the 4-velocity are well defined.
The gauge degrees of freedom in the choice of frame are analysed in Appendix 
\ref{Frame}. Applying the commutation relations given in Appendix \ref
{commutation} and the useful relations for the vector and tensor spherical
harmonics given in Appendix \ref{harmonics}, the equations governing the
perturbed system are derived in Appendix \ref{perturbEq_harmonics}. Then by
fixing a frame we find that the perturbed spacetime can be described by six
type of harmonic coefficients. The evolution equations for these six
variables are given in Section \ref{perturb_sec}. The behaviour of the
perturbations in a geometrical optics approximation is discussed in Section 
\ref{GeomOp}, while Section \ref{Concl} contains some
concluding remarks.

Units where $8\pi G=1$ and $c=1$ are used throughout this paper.


\section{The 1+3 and 1+1+2 covariant formalisms\label{decomp_sec}}


\subsection{The 1+3 covariant formalism}

Let $u^{a}$ be a time-like vector field obeying the usual normalisation
condition $u^{a}u_{a}=-1$ and $h_{ab}$ a spatial 3-metric satisfying $
u^{a}h_{ab}=0$. Then the 4-metric $g_{ab}$\ can be decomposed as 
\begin{equation}
g_{ab}=-u_{a}u_{b}+h_{ab}\ .
\end{equation}
We denote the 4-dimensional (4D) and 3-dimensional (3D) volume elements by $
\eta _{abcd}=\sqrt{-g}\delta _{\,\,[a}^{0}\delta _{\,\,b}^{1}\delta
_{\,\,c}^{2}\delta _{\,\,d]}^{3}$ and $\varepsilon _{abc}=\eta _{dabc}u^{d}$%
, respectively. Angular brackets $\langle ~\rangle $ on indices denote
symmetrised and trace-free tensors which are projected in all indices with
the metric $h_{ab}$. Round brackets $(~)$ and square brackets $[~]$ on
indices denote the symmetric and antisymmetric parts, respectively. A dot
denotes covariant derivatives along the integral curves of $u^{a}$, while $%
D_a$ is the projected spatial derivative 
\begin{eqnarray}
\dot{T}_{b..c} &=&u^{a}\nabla _{a}T_{b..c}~, \\
D_{a}T_{b..c} &=&h_{a}^{\,\,\,d}h_{b}^{\,\,\,i}..h_{c}^{\,\,\,\,j}\nabla
_{d}T_{i..j}\ .
\end{eqnarray}
For vanishing vorticity of $u^a$, $D_a$ is the 3D covariant derivative
compatible with the metric $h_{ab}$.

The kinematic quantities are introduced through the decomposition of the 4D
covariant derivative of $u^{a}$ as%
\begin{equation}
\nabla _{a}u_{b}=\sigma _{ab}+\frac{1}{3}\Theta h_{ab}+\omega
_{ab}-u_{a}A_{b}\ ,  \label{1+3Decomp}
\end{equation}%
where $\sigma _{ab}=D_{\langle a}u_{b\rangle }$ is the shear, $\Theta
=D^{a}u_{a}$ the expansion and $\omega _{ab}=D_{[a}u_{b]}$ the vorticity of $%
u^{a}$, finally $A_{a}=\dot{u}_{\langle a\rangle }%
=h_a{}^{\;b}\dot u_b$ is its acceleration. Since $\omega _{ab}$ is space-like and
antisymmetric, containing 3 independent components, we introduce its
(Hodge-) dual $\omega _{a}=\varepsilon _{a}^{\,\,\,\,bc}\omega _{bc}/2$.

The gravito-electro-magnetic quantities arise from the 1+3 covariant
decomposition of the 4D Weyl tensor $C_{abcd}$ as 

\begin{equation}
E_{ab}=C_{acbd}u^{c}u^{d} \quad
\hbox{and} \quad  H_{ab}=\frac{1}{2}\varepsilon _{a}^{\,\,\,\,\,\,cd}C_{cdbe}u^{e}. 
\end{equation}
The quantities $E_{ab}$ and $H_{ab}$ are the magnetic and
electric parts of $C_{abcd}$, respectively. 
The Weyl tensor is given\footnote{The definition of $E_{ab}$ differs by a sign in \cite{Hawking}.} \cite{Hawking} by 
\begin{equation}
\frac{1}{2}C_{abcd}=u_{c}u_{[a}E_{b]d}-u_{d}u_{[a}E_{b]c}+E_{c[a}h_{b]d}-E_{d[a}h_{b]c}-\varepsilon _{ab}^{\ \ \ \ i}H_{i[c}u_{d]}-\varepsilon _{cd}^{\ \ \ \ i}H_{i[a}u_{b]}~.
\label{Weyldecomp}
\end{equation}

As well-known, the Weyl tensor (posessing the symmetries of the Riemann
tensor with 20 independent components, and also being traceless, meaning 10
conditions) has 10 independent components. Its electric and magnetic parts
have each 5 independent components, as they are expressed by traceless and
symmetric 3-tensors.

The energy-momentum tensor $T_{ab}$ of matter fields is decomposed with
respect to an observer with 4-velocity $u^{a}$ in the standard way: 
\begin{equation}
T_{ab}=\mu u_{a}u_{b}+2q_{(a}u_{b)}+ph_{ab}+\pi _{ab}\ .  \label{EnMomTens}
\end{equation}%
The quantities $\mu $, $q_{a}$, $p$ and $\pi _{ab}$ are the energy density,
the energy current vector, the isotropic pressure and the symmetric,
trace-free anisotropic pressure tensor of matter.

The Riemann tensor can be expressed by the metric components,
gravito-electro-magnetic quantities and matter variables as follow. First
we use the decomposition of $R_{abcd}$ into its Weyl and Ricci ($R_{ab}$)
contributions
\begin{equation}
R_{abcd}=C_{abcd}+g_{a[c}R_{d]b}-g_{b[c}R_{d]a}-\frac{R}{3}g_{a[c}g_{d]b}~,
\end{equation}
where $R$ is the Ricci scalar. Then by applying the Einstein equation\begin{equation}
R_{ab}=\Lambda g_{ab}+T_{ab}-\frac{T}{2}g_{ab}~,  \label{EinEq}
\end{equation}with $T=g^{ab}T_{ab}$ and cosmological constant $\Lambda $, and Eqs. (\ref{Weyldecomp})-(\ref{EnMomTens}), we find
\begin{eqnarray}
R_{abcd} &=&2\left(
u_{c}u_{[a}E_{b]d}-u_{d}u_{[a}E_{b]c}+E_{c[a}h_{b]d}-E_{d[a}h_{b]c}-\varepsilon _{ab}^{\ \ \ \ i}H_{i[c}u_{d]}-\varepsilon _{cd}^{\ \ \ \
i}H_{i[a}u_{b]}\right)  \nonumber \\
&&-\frac{\left( 2\Lambda -\mu -3p\right) }{3}\left(
u_{a}u_{[c}h_{d]b}-u_{b}u_{[c}h_{d]a}\right) +\frac{2\left( \Lambda +\mu
\right) }{3}h_{a[c}h_{d]b}  \nonumber \\
&&+q_{a}u_{[c}h_{d]b}-q_{b}u_{[c}h_{d]a}+u_{a}q_{[c}h_{d]b}-u_{b}q_{[c}h_{d]a}+g_{a[c}\pi _{d]b}-g_{b[c}\pi _{d]a}~.
\label{4DRiemannexpr}
\end{eqnarray}

The full set of equations arise from the Ricci identities for $u^{a}$ and
from the 4D Bianchi identities and can be found for instance in Refs. \cite%
{Cargese}, \cite{Bonometto}.

A 3D curvature tensor can be defined in the following way (see \cite{Tsagas}%
):%
\begin{equation}
\frac{1}{2}^{\left( 3\right) }R_{abcd}V^{d}=\textcolor{black}{D _{\lbrack a}D
_{b]}}V_{c}-\omega _{ab}\dot{V}_{\langle c\rangle }\ ,  \label{3dCurvTens}
\end{equation}%
where $V^{a}$ is an arbitrary 3-vector. In the vorticity-free case {%
\color{black} (as a consequence of Frobenius's theorem)} $^{\left( 3\right)
}R_{abcd}$ is the Riemann curvature of the hypersurface with metric $h_{ab}$%
. 
\textcolor{black}{Alternatively it can be given in terms of the Gauss' equation \textcolor{black}{\cite{HawkingEllis}:}
\begin{equation}
^{\left( 3\right) }R_{abcd}=h_a^{\;e}h_b^{\;f}h_c^{\;g}h_d^{\;h}
\; ^{\left( 4\right) }R_{abcd}-(D_c u_a) (D_d u_b)+(D_d u_a) (D_c u_b) \, . \label{Gauss}
\end{equation}
From Eqs. (\ref{Gauss}), (\ref{1+3Decomp}) and (\ref{EinEq})} the curvature
scalar $^{\left( 3\right) }R=h^{ac}h^{bd\left( 3\right) }R_{abcd}$ is %
\textcolor{black}{\cite{Tsagas}:}%
\begin{equation}
\frac{^{\left( 3\right) }R}{2}=\ \mu +\Lambda -\frac{\Theta ^{2}}{3}+\sigma
^{ab}\sigma _{ab}-\omega _{ab}\omega ^{ab}  \label{3dCurvScal}
\end{equation}%
giving the usual Ricci scalar on a spatial hypersurface orthogonal to $u^a$
(when $\omega _{ab}=0$).


\subsection{The 1+1+2 covariant formalism}

When there is a unique preferred spatial direction at each point, it is
worthwhile performing a further decomposition of spacetime into its so
called 1+1+2 form. The preferred spatial direction will be singled out by a
normalised vector field $n^{a}$ ($n^{a}n_{a}=1\;,n^{a}u_{a}=0$) and this
allows us to decompose the metric $h_{ab}$ as%
\begin{equation}
h_{ab}=n_{a}n_{b}+N_{ab}\ ,  \label{hdec}
\end{equation}%
where $N_{ab}$ is the induced 2-metric on the surface perpendicular to both $%
n^{a}$ and $u^{a}$. The alternating Levi-Civita 2-tensor is defined as $%
\varepsilon _{ab}=\varepsilon _{abc}n^{c}$. Curly brackets $\{$ $\}$ on
indices will denote symmetrised and trace-free tensors which are projected
in all indices with the metric $N_{ab}$. A bar on vector indices will denote
projection onto the 2-sphere: $v_{\bar{a}}\equiv N_{ab}v^{b}$. The 3D
covariant derivative can then be projected into two parts as \cite{1+1+2}%
\begin{eqnarray}
\widehat{T}_{b..c} &=&n^{a}D_{a}T_{b..c}~,  \label{hatderivative} \\
\delta _{a}T_{b..c}
&=&N_{a}^{\,\,\,d}N_{b}^{\,\,\,i}..N_{c}^{\,\,\,\,j}D_{d}T_{i..j}\ .
\end{eqnarray}%
A further decomposition of the 1+3 vector and tensor variables with respect
to $n^{a}$ are given according to Ref. \cite{1+1+2}. The kinematical vectors
and tensor can be decomposed as%
\begin{equation}
A^{a}=\mathcal{A}n^{a}+\mathcal{A}^{a}\ ,  \label{A3dec}
\end{equation}%
\begin{equation}
\omega ^{a}=\Omega n^{a}+\Omega ^{a}\ ,  \label{om3dec}
\end{equation}%
\begin{equation}
\sigma _{ab}=\Sigma \left( n_{a}n_{b}-\frac{1}{2}N_{ab}\right) +2\Sigma
_{(a}n_{b)}+\Sigma _{ab}\ ,  \label{sigma3dec}
\end{equation}%
while the gravito-electro-magnetic variables are%
\begin{equation}
E_{ab}=\mathcal{E}\left( n_{a}n_{b}-\frac{1}{2}N_{ab}\right) +2\mathcal{E}%
_{(a}n_{b)}+\mathcal{E}_{ab}\ ,
\end{equation}%
\begin{equation}
H_{ab}=\mathcal{H}\left( n_{a}n_{b}-\frac{1}{2}N_{ab}\right) +2\mathcal{H}%
_{(a}n_{b)}+\mathcal{H}_{ab}\ .
\end{equation}%
\textcolor{black}{In the above decompositions the scalars, the 2-vectors and the
symmetric 2-tensors represent one, two and two independent components each,
respectively. }Finally, the 1+2 form of the energy current vector and
anisotropic pressure tensor are%
\begin{equation}
q_{a}=Qn_{a}+Q_{a}\ ,
\end{equation}%
\begin{equation}
\pi _{ab}=\Pi \left( n_{a}n_{b}-\frac{1}{2}N_{ab}\right) +2\Pi
_{(a}n_{b)}+\Pi _{ab}\ .
\end{equation}%
Here $\mathcal{A}^{a}$, $\Omega ^{a}$, $\Sigma ^{a}$, $\mathcal{E}^{a}$, $%
\mathcal{H}^{a}$, $Q^{a}$ and $\Pi ^{a}$ are 2-vectors and $\Sigma _{ab}$, $%
\mathcal{E}_{ab}$, $\mathcal{H}_{ab}$ and $\Pi _{ab}$ are trace-free,
symmetric 2-tensors perpendicular to both $u^{a}$ and $n^{a}$.

The additional fundamental variables of the 1+1+2 formalism arise from the
projected time derivative of $n^{a}$ 
\textcolor{black}{and the 3D covariant derivative of $n_a$. The time derivative can be written as $\dot{n}_{a}=\mathcal{B}u_{a}+\alpha _{a}$, where  $\alpha _{a}$ is a 2-vector (since $\dot{n}_{a}n^a=0$). Now, from $u^a n_a=0$, it follows that $u^a \dot n_a= -n_a A^a$. Using (\ref{A3dec}) one obtains $\mathcal{B}=\mathcal{A}$.
Hence 
\begin{equation}
\dot{n}_{a}=\mathcal{A}u_{a}+\alpha _{a}
\end{equation}
($\alpha _{a}=\dot{n}_{\bar{a}} $).
The 3D covariant derivative of $n_{a}$ can be decomposed as:}%
\begin{equation}
D_{a}n_{b}=\zeta _{ab}+\frac{\phi }{2}N_{ab}+\xi \varepsilon
_{ab}+n_{a}a_{b}\ ,  \label{Dndec}
\end{equation}%
where $\phi $ is the sheet expansion, $a_{a}=n^{c}D_{c}n_{a}$ is the
acceleration, $\varsigma _{ab}$ is the shear of $n^{a}$ and $\xi $
represents its rotation in the local 3D space. Here $\alpha _{a}$ and $a_{a}$
are 2-vectors and $\zeta _{ab}$ is a trace-free, symmetric 2-tensor
perpendicular to both $u^{a}$ and $n^{a}$.

The full set of evolution and constraint equations for perturbed LRS
spacetimes are given in Ref. \cite{1+1+2}. There are no evolution equations
for $\mathcal{A}$, $\mathcal{A}_{a}$, $\alpha _{a}$ and there is no
propagation equation for $a_{a}$. These are determined by fixing a
particular frame \cite{1+1+2}.

We define the 2-dimensional (2D) curvature tensor $\mathcal{R}_{abcd}$ as%
\begin{equation}
\frac{1}{2}\mathcal{R}_{abcd}V^{d}=\delta _{\lbrack a}\delta
_{b]}V_{c}-\Omega \varepsilon _{ab}\dot{V}_{\bar{c}}+\xi \varepsilon _{ab}%
\widehat{V}_{\bar{c}}\ ,  \label{2Dcurvtensdef}
\end{equation}%
where $V^{a}$ is an arbitrary 2-vector. Similar definitions are given in 
\cite{3+1+1} for higher dimensional spacetimes. For vanishing $\Omega $ and $%
\xi $ it agrees with the usual Riemann curvature tensor of $N_{ab}$. This
definition gives (\textcolor{black}{see Appendix \ref{2Dcurvtens}})%
\begin{equation}
\mathcal{R}_{abcd}=N_{a}^{i}N_{b}^{j}N_{c}^{k}N_{d}^{l}R_{ijkl}+\left(
\delta _{a}u_{d}\right) \left( \delta _{b}u_{c}\right) -\left( \delta
_{a}u_{c}\right) \left( \delta _{b}u_{d}\right) -\left( \delta
_{a}n_{d}\right) \left( \delta _{b}n_{c}\right) +\left( \delta
_{a}n_{c}\right) \left( \delta _{b}n_{d}\right) \ ,  \label{2Dcurvtensdef2}
\end{equation}%
where $R_{ijkl}$ is the usual 4D Riemann tensor. 
\textcolor{black}{By using Eqs. (\ref{4DRiemannexpr}), (\ref{1+3Decomp}) and (\ref{A3dec})-(\ref{Dndec}), $\mathcal{R}_{abcd}$ is expressed as\begin{eqnarray}
\mathcal{R}_{abcd} &=&\left[ \frac{2\left( \Lambda +\mu \right) }{3}-\Pi -2\mathcal{E}-\frac{1}{2}\left( \Sigma -\frac{2\Theta }{3}\right) ^{2}+\frac{\phi ^{2}}{2}\right] N_{c[a}N_{b]d}  \nonumber \\
&&+N_{c[a}\left\{ \Pi _{b]d}+2\mathcal{E}_{b]d}+\left( \Sigma -\frac{2\Theta 
}{3}\right) \Sigma _{b]d}+\phi \zeta _{b]d}+\left[ \Omega \left( \Sigma -\frac{2\Theta }{3}\right) +\xi \phi \right] \varepsilon _{b]d}\right\} 
\nonumber \\
&&-N_{d[a}\left\{ 2\Pi _{b]c}+2\mathcal{E}_{b]c}+\left( \Sigma -\frac{2\Theta }{3}\right) \Sigma _{b]c}+\phi \zeta _{b]c}+\left[ \Omega \left(
\Sigma -\frac{2\Theta }{3}\right) +\xi \phi \right] \varepsilon
_{b]c}\right\}  \nonumber \\
&&+2\varepsilon _{c[a}\left[ \Omega \Sigma _{b]d}-\xi \zeta _{b]d}\right]
-2\varepsilon _{d[a}\left[ \Omega \Sigma _{b]c}-\xi \zeta _{b]c}\right] 
\nonumber \\
&&+2\left( \Omega ^{2}-\xi ^{2}\right) \varepsilon _{c[a}\varepsilon
_{b]d}+2\left( \zeta _{c[a}\zeta _{b]d}-\Sigma _{c[a}\Sigma _{b]d}\right) ~.
\end{eqnarray}} Defining the 2D curvature scalar $\mathcal{R}=N^{ac}N^{bd}$~$%
\mathcal{R}_{abcd}$, we find%
\begin{equation}
\mathcal{R}=\frac{2}{3}\left( \mu +\Lambda \right) -\Pi -2\mathcal{E}-\frac{1%
}{2}\left( \Sigma -\frac{2\Theta }{3}\right) ^{2}+\frac{\phi ^{2}}{2}%
-2\left( \Omega ^{2}-\xi ^{2}\right) +\Sigma _{ab}\Sigma ^{ab}-\zeta
_{ab}\zeta ^{ab}\ .  \label{2dcurvscalar}
\end{equation}%
%

\section{The Kantowski-Sachs background\label{KSbackqround_sec}}

\textcolor{black}{The spatial sections of Kantowski-Sachs cosmologies have topology $R\times S^2$  
and are the only spatially homogeneous cosmologies that do not fit into the Bianchi classification. This is due to that their isometry group does not admit a 3-dimensional subgroup that acts simply transitive on the hypersurfaces of homogenity. These metrics are Locally Rotationally Symmetric (LRS) and  belong to the LRS class II, characterised by $\omega_{ab}=\xi=H_{ab}=0$
\cite{LRS,MarklundBradley}.} 
\textcolor{black}{T}he square of the line-element can be written as%
\begin{equation}
ds^{2}=-dt^{2}+a_{1}^{2}\left( t\right) dz^{2}+a_{2}^{2}\left( t\right)
\left( d\vartheta ^{2}+\sin ^{2}\vartheta d\varphi ^{2}\right) \ .
\label{KASA}
\end{equation}%
The 4-velocity of comoving observers is $u=\partial /\partial t$ and the
direction of anisotropy is $n=a_{1}^{-1}\partial /\partial z$, where $z$ is
dimensionless. The coordinates $\vartheta $ and $\varphi $ are the polar and
azimuthal angles on $S^{2}$, respectively. The scales $a_{1}$ and $a_{2}$
have the dimension of time, $a_{2}$ being assumed to be sufficiently large
to avoid periodic structures in the angles emerging in the observable
Universe. Symmetry and normalisation implies \cite{LRS}:%
\begin{equation}
\hat{n}_{b}\equiv n^{a}D_{a}n_{b}=0\ \ ,\ \ \ \dot{n}_{\bar{b}}=0\ \ ,
\end{equation}%
i.e., $n_{a}$ is geodesic on local 3-space with metric $h_{ab}$ and is Fermi
propagated along the integral curves of $u^{a}$. In the spacetime given by
Eq. (\ref{KASA}) the non-vanishing kinematical variables of 1+1+2 formalism
are the expansion \cite{KASAscalar}:%
\begin{equation}
\Theta =\frac{\dot{a}_{1}}{a_{1}}+2\frac{\dot{a}_{2}}{a_{2}}\ ,
\label{Thetabackgroundexp}
\end{equation}%
and the scalar part 
\begin{equation}
\Sigma =\frac{2}{3}\left( \frac{\dot{a}_{1}}{a_{1}}-\frac{\dot{a}_{2}}{a_{2}}%
\right)  \label{Sigmabackgroundexp}
\end{equation}%
of the shear $\sigma _{ab}$. Given an equation of state $p=p\left( \mu
\right) $ for the pressure $p$ and energy density $\mu $ of the perfect
fluid, for any given cosmological constant, the Kantowski-Sachs models are
completely determined in terms of the shear $\Sigma $, expansion $\Theta $
and $\mu $. The electric part of 4D Weyl tensor is then determined
algebraically as (see Eq. (100) in \cite{1+1+2})%
\begin{equation}
3\mathcal{E}=-2\left( \mu +\Lambda \right) -3\Sigma ^{2}+\frac{2}{3}\Theta
^{2}+\Sigma \Theta \ .  \label{EWeyl}
\end{equation}

The evolutions of $\Sigma $, $\Theta $ and $\mu $ are governed by Eqs. (96),
(94) and (95) of \cite{1+1+2}:%
\begin{equation}
\dot{\mu}=-\Theta \left( \mu +p\right) \ ,  \label{continuity}
\end{equation}%
\begin{equation}
\dot{\Theta}=-\frac{\Theta ^{2}}{3}-\frac{3}{2}\Sigma ^{2}-\frac{1}{2}\left(
\mu +3p\right) +\Lambda \ ,  \label{expansiondot}
\end{equation}%
\begin{equation}
\dot{\Sigma}=\frac{2}{3}\left( \mu +\Lambda \right) +\frac{\Sigma ^{2}}{2}%
-\Sigma \Theta -\frac{2}{9}\Theta ^{2}\ ,  \label{Sigmadot}
\end{equation}%
where we have used Eq. (\ref{EWeyl}).

For Kantowski-Sachs spacetime the 2D scalar curvature (\ref{2dcurvscalar})
becomes%
\begin{equation}
\mathcal{R}=\frac{2}{3}\left( \mu +\Lambda \right) -2\mathcal{E}-\frac{1}{2}%
\left( \Sigma -\frac{2\Theta }{3}\right) ^{2}=2\left( \mu +\Lambda \right) +%
\frac{3}{2}\Sigma ^{2}-\frac{2\Theta ^{2}}{3}=\frac{2}{a_{2}^{2}}\ ,
\label{2TimesGaussianCurv}
\end{equation}%
that is two times the Gaussian curvature of the 2-spheres. Taking the time
derivative of $\mathcal{R}$, and using Eqs. (\ref{continuity})-(\ref%
{Sigmadot}), we find 
\begin{equation}
\mathcal{\dot{R}}=\left( \Sigma -\frac{2\Theta }{3}\right) \mathcal{R}\ .
\label{Kevo}
\end{equation}%
One of the evolution equations (\ref{continuity})-(\ref{Sigmadot}) can be
replaced by (\ref{Kevo}).

In summary the non-vanishing quantities on the background are given by the
set 
\begin{equation}
S^{(0)}\equiv \{\Theta ,\Sigma ,{\mathcal{E}},\mu ,p\}
\end{equation}%
or equivalently 
\begin{equation}
S^{(0)}=\{\Theta ,\Sigma ,{\mathcal{R}},\mu ,p\}\,.
\end{equation}%
To zeroth order (on the background) $\mathcal{E}$ (or $\mathcal{R}$) are
given in terms of the other quantities.

General orthogonal spatially homogeneous LRS class II spacetimes emerge as
slight modifications \cite{MarklundBradley}. The Kantowski-Sachs cosmologies
are the only ones with $\mathcal{R}>0$. If $\mathcal{R}<0$ the spacetimes
are of Bianchi type III and the only modifications to the above equations
are to replace $\sin \vartheta $ by $\sinh \vartheta $ in equation (\ref%
{KASA}) and to change $\mathcal{R}$ to $\mathcal{R}=-2/a_{2}^{2}$. For $%
\mathcal{R}=0$ there are solutions of Bianchi type I/VII$_{0}$. Due to that (%
\ref{2TimesGaussianCurv}) now becomes a (satisfied) constraint, one of the
evolution equations can be dropped. There are also ${\mathcal{R}}=0$ with $%
\Sigma ={\mathcal{E}}=0$. For these the sheet expansion $\phi $ is in
general nonzero and the system is given by equations (\ref{continuity}) and (%
\ref{expansiondot}) plus the constraint 
\begin{equation}
\mu +\Lambda -\frac{1}{3}\Theta ^{2}+\frac{3}{4}\phi ^{2}=0\,.
\end{equation}%
\textcolor{black}{These are flat (when $\phi =0$) or negatively curved
Friedmann models, which also fall into the Bianchi I and V classes respectively.}

In this paper we consider perturbations of the ${\mathcal{R}}>0$ models, but
the above discussion shows that perturbations of the ${\mathcal{R}}<0$ can
be done in an analogous way. Perturbations of Friedmann models and Bianchi I
models have been considered elsewhere \cite{cov1, cov2, cov3, cov4, cov5,
cov6, cov7}.


\section{Vorticity-free, perfect fluid perturbations of Kantowski-Sachs
cosmologies\label{perturb_sec}}


{\color{black} For simplicity,} we assume the perturbed fluid is irrotational
and we use a frame associated with the fluid. This requirement assigns the
reference 4-velocity in the perturbed spacetime and in a 1+3 covariant
formalism the frame is completely fixed. In this frame the energy current $%
q_{a}$ and $\omega ^{a}=\Omega n^{a}+\Omega ^{a}$ vanish. Moreover, we
neglect the anisotropic pressure contributions to the energy-momentum
tensor, i.e., $\pi _{ab}=0$, {\color{black} restricting our analysis to
barotropic perfect fluids}.

The variables of 1+1+2 formalism are defined with respect to the frame
vectors $u$ and $n$. Therefore the variables are not frame-invariant in
general (see Appendix \ref{Frame}). (Of course their combinations may result
in frame-invariant quantities \cite{Schperturb}.) The frame choice does not
fix completely the mapping between the perturbed and the background geometry 
\cite{cov1}, \cite{Brunietal}, \cite{BruniSonego}. The variables vanishing
on the background are invariant for the remaining gauge fixing in this map
according to the Stewart-Walker lemma \cite{StewartWalker}. Therefore
instead of $\mathcal{E}$, $\Theta $, $\Sigma $, $\mu $ and $p$ we use their
gradients%
\begin{eqnarray}
X_{a} &=&\delta _{a}\mathcal{E}\ ,\ V_{a}=\delta _{a}\Sigma \ ,\
W_{a}=\delta _{a}\Theta \ ,  \nonumber \\
\mu _{a} &=&\delta _{a}\mu \ ,\ p_{a}=\delta _{a}p\ ,  \label{gaugeinv}
\end{eqnarray}%
that vanish on the background. As will be shown in section \ref%
{harmonicrelations}, the hat-derivatives (\ref{hatderivative}) are related
to the $\delta _{a}$ derivatives when the vorticity vanishes.

Hence we have the following nonzero first order quantities (that vanish on
the background): 
\begin{equation}
S^{(1)}\equiv \left\{ X_{a},V_{a},W_{a},\mu _{a},p_{a},{\mathcal{A}},{%
\mathcal{A}}_{a},\Sigma _{a},\Sigma _{ab},{\mathcal{E}}_{a},{\mathcal{E}}%
_{ab},{\mathcal{H}}_{a},{\mathcal{H}}_{ab},a_{b},\phi ,\xi ,\zeta
_{ab}\right\} ~.
\end{equation}

From the 1+1+2 equations in \cite{1+1+2} the following nontrivial evolution
equations then hold on the perturbed Kantowski-Sachs spacetime \footnote{%
There are some minor misprints in \cite{1+1+2}. In Eq. (36) the term $%
(\Sigma _{a}-\epsilon _{ab})\Omega ^{b}\dot{\psi}$ should probably read $%
-2\epsilon _{ab}\Omega ^{b}\dot{\psi}$, in Eq. (40) $(\Sigma _{a}-\epsilon
_{ac})\Omega ^{c}\dot{\psi}_{\bar{b}}\rightarrow -2\epsilon _{ac}\Omega ^{c}%
\dot{\psi}_{\bar{b}}$, in Eq. (52) the term $-(\frac{2}{3}\Theta +\frac{1}{2}%
\Sigma )\Sigma _{ab}\rightarrow -(\frac{2}{3}\Theta -\Sigma )\Sigma _{ab}$
in Eq. (53) the terms $-\xi \epsilon _{ab}\alpha ^{b}+(\frac{1}{3}\Theta
+\Sigma )({\mathcal{A}}_{a}-a_{a})\rightarrow +\xi \epsilon _{ab}\alpha
^{b}+(\frac{1}{3}\Theta +\Sigma )({\mathcal{A}}_{a}+a_{a})$, in Eq. (76) $%
-\Sigma \epsilon _{ab}{\mathcal{H}}^{b}\rightarrow -\frac{3}{2}\Sigma
\epsilon _{ab}{\mathcal{H}}^{b}$ and in Eq. (80) $-(\frac{1}{3}\Theta -\frac{%
1}{2}\Sigma )(\Sigma _{a}-\epsilon _{ab}\Omega ^{b})\rightarrow +(\frac{1}{3}%
\Theta -\frac{1}{2}\Sigma )(\Sigma _{a}-\epsilon _{ab}\Omega ^{b})$.}:%
\begin{equation}
\dot{\phi}=\left( \Sigma -\frac{2\Theta }{3}\right) \left( \frac{\phi }{2}-%
\mathcal{A}\right) +\delta ^{a}\alpha _{a}\ ,  \label{phidot}
\end{equation}%
\begin{equation}
2\dot{\xi}=\left( \Sigma -\frac{2\Theta }{3}\right) \xi +\varepsilon
^{ab}\delta _{a}\alpha _{b}+\mathcal{H}\ ,
\end{equation}%
\begin{equation}
\dot{\mathcal{H}}=\frac{3}{2}\!\!\left( \Sigma -\frac{2\Theta }{3}\right) 
\mathcal{H}-\varepsilon ^{ab}\delta _{a}\mathcal{E}_{b}-3\mathcal{E}\xi \ ,
\end{equation}%
\begin{equation}
\dot{\mu}_{\bar{a}}=\frac{1}{2}\left( \Sigma -\frac{2\Theta }{3}\right) \mu
_{a}-\Theta \left( \mu _{a}+p_{a}\right) -\left( \mu +p\right) W_{a}+\dot{\mu%
}\mathcal{A}_{a}\ ,  \label{odden}
\end{equation}%
\begin{equation}
\dot{X}_{\bar{a}}=2\!\left( \!\Sigma \!-\!\frac{2\Theta }{3}\!\right)
\!X_{a}\!+\!\frac{3\mathcal{E}}{2}\!\left( \!V_{a}\!\!-\!\frac{2}{3}%
W_{a}\!\right) \!-\!\frac{\mu +p}{2}V_{a}-\frac{\Sigma }{2}\left( \mu
_{a}+p_{a}\right) +\dot{\mathcal{E}}\mathcal{A}_{a}+\varepsilon ^{bc}\delta
_{a}\delta _{b}\mathcal{H}_{c}\ ,
\end{equation}%
\begin{equation}
\dot{V}_{\bar{a}}\!-\!\frac{2}{3}\dot{W}_{\bar{a}}\!=\frac{3}{2}\left(
\Sigma -\frac{2\Theta }{3}\right) \!\left( V_{a}-\frac{2}{3}W_{a}\right)
-X_{a}+\frac{1}{3}\left( \mu _{a}+3p_{a}\right) +\!\left( \!\dot{\Sigma}\!-\!%
\frac{2\dot{\Theta}}{3}\!\right) \!\mathcal{A}_{a}-\delta _{a}\delta ^{b}%
\mathcal{A}_{b},
\end{equation}%
\begin{equation}
\dot{\Sigma}_{\{ab\}}=\left( \Sigma -\frac{2\Theta }{3}\right) \Sigma
_{ab}+\delta _{\{a}\mathcal{A}_{b\}}-\mathcal{E}_{ab}\ ,
\end{equation}%
\begin{equation}
\dot{\zeta}_{\{ab\}}=\frac{1}{2}\left( \Sigma -\frac{2\Theta }{3}\right)
\zeta _{ab}+\delta _{\{a}\alpha _{b\}}-\varepsilon _{c\{a}\mathcal{H}%
_{b\}}^{\,\,\,\,\,c}\ .
\end{equation}%
The equations containing both propagation and evolution contributions are%
\begin{equation}
\dot{W}_{\bar{a}}-\delta _{a}\widehat{\mathcal{A}}=\left( \frac{\Sigma }{2}%
-\Theta \right) W_{a}-3\Sigma V_{a}-\frac{1}{2}\left( \mu _{a}+3p_{a}\right)
+\dot{\Theta}\mathcal{A}_{a}+\delta _{a}\delta ^{b}\mathcal{A}_{b}\ ,
\end{equation}%
\begin{equation}
\widehat{\alpha }_{\bar{a}}-\dot{a}_{\bar{a}}=\left( \Sigma +\frac{\Theta }{3%
}\right) \left( \mathcal{A}_{a}+a_{a}\right) -\varepsilon _{ab}\mathcal{H}%
^{b}\ ,
\end{equation}%
\begin{equation}
2\dot{\Sigma}_{\bar{a}}-\widehat{\mathcal{A}}_{a}=\delta _{a}\mathcal{A}%
-\left( \Sigma +\frac{4\Theta }{3}\right) \Sigma _{a}-3\Sigma \alpha _{a}-2%
\mathcal{E}_{a}\ ,
\end{equation}%
\begin{equation}
\dot{\mathcal{E}}_{\bar{a}}\!+\frac{1}{2}\varepsilon _{ab}\widehat{\mathcal{H%
}}^{b}=\!\frac{3}{4}\left( \!\Sigma \!-\frac{4\Theta }{3}\!\right) \!%
\mathcal{E}_{a}+\frac{\left( 3\mathcal{E\!}-\!2\mu \!-\!2p\right) }{4}\Sigma
_{a}+\frac{3}{4}\varepsilon _{ab}\delta ^{b}\mathcal{H\!}\!-\frac{3\mathcal{E%
}}{2}\alpha _{a}\!+\frac{1}{2}\varepsilon _{bc}\delta ^{b}\mathcal{H}%
_{\,\,a}^{c},
\end{equation}%
\begin{equation}
\dot{\mathcal{H}}_{\bar{a}}\!-\frac{1}{2}\varepsilon _{ab}\widehat{\mathcal{E%
}}^{b}=\frac{3}{4}\left( \Sigma -\frac{4\Theta }{3}\right) \mathcal{H}_{a}-%
\frac{3}{4}\varepsilon _{ab}X^{b}-\frac{3\mathcal{E}}{2}\varepsilon _{ab}%
\mathcal{A}^{b}\!+\frac{3\mathcal{E}}{4}\varepsilon _{ab}a^{b}\!-\frac{1}{2}%
\varepsilon _{bc}\delta ^{b}\mathcal{E}_{\,\,a}^{c},
\end{equation}%
\begin{equation}
\dot{\mathcal{E}}_{\{ab\}}\!-\varepsilon _{c\{a}\widehat{\mathcal{H}}%
_{b\}}^{\,\,\,\,\,c}\!=-\frac{3}{2}\!\left( \Sigma \!+\!\frac{2\Theta }{3}%
\right) \!\mathcal{E}_{ab}\!-\!\varepsilon _{c\{a}\delta ^{c}\mathcal{H}%
_{b\}}-\frac{\left( 3\mathcal{E}+\mu +p\right) }{2}\Sigma _{ab},
\end{equation}%
\begin{equation}
\dot{\mathcal{H}}_{\{ab\}}+\varepsilon _{c\{a}\widehat{\mathcal{E}}%
_{b\}}^{\,\,\,\,\,c}=-\frac{3}{2}\left( \Sigma +\frac{2\Theta }{3}\right) 
\mathcal{H}_{ab}+\frac{3\mathcal{E}}{2}\varepsilon _{c\{a}\zeta
_{b\}}^{\,\,\,\,c}+\varepsilon _{c\{a}\delta ^{c}\mathcal{E}_{b\}}\ .
\end{equation}%
The pure propagation equations are%
\begin{equation}
\widehat{\phi }=-\left( \Sigma -\frac{2\Theta }{3}\right) \left( \Sigma +%
\frac{\Theta }{3}\right) +\delta ^{a}a_{a}-\mathcal{E}-\frac{2\left( \mu
+\Lambda \right) }{3},
\end{equation}%
\begin{equation}
2\widehat{\xi }=\varepsilon ^{ab}\delta _{a}a_{b}\ ,
\end{equation}%
\begin{equation}
\widehat{\mathcal{H}}=-\delta ^{a}\mathcal{H}_{a}\ ,
\end{equation}%
\begin{equation}
\widehat{\mathcal{A}}_{a}=\delta _{a}\mathcal{A}\ ,
\end{equation}%
\begin{equation}
\widehat{p}_{\bar{a}}=-\left( \mu +p\right) \delta _{a}\mathcal{A}\ ,
\end{equation}%
\begin{equation}
\frac{\widehat{\mu }_{\bar{a}}}{3}-\widehat{X}_{\bar{a}}=\frac{3\mathcal{E}}{%
2}\delta _{a}\phi +\delta _{a}\delta ^{b}\mathcal{E}_{b}\!\!\ ,
\end{equation}%
\begin{equation}
\frac{2}{3}\widehat{W}_{\bar{a}}-\widehat{V}_{\bar{a}}=\frac{3\Sigma }{2}%
\delta _{a}\phi +\delta _{a}\delta ^{b}\Sigma _{b}\ ,
\end{equation}%
\begin{equation}
\widehat{\Sigma }_{\bar{a}}=\frac{1}{2}\left( V_{a}+\frac{4}{3}W_{a}\right) -%
\frac{3\Sigma }{2}a_{a}-\delta ^{b}\Sigma _{ab}\ ,
\end{equation}%
\begin{equation}
2\widehat{\mathcal{E}}_{\bar{a}}=X_{a}-3\mathcal{E}a_{a}-3\Sigma \varepsilon
_{ab}\mathcal{H}^{b}-2\delta ^{b}\mathcal{E}_{ab}\!+\frac{2}{3}\mu _{a}\ ,
\end{equation}%
\begin{equation}
2\widehat{\mathcal{H}}_{\bar{a}}=\delta _{a}\mathcal{H}-3\mathcal{E}%
\varepsilon _{ab}\Sigma ^{b}+3\Sigma \varepsilon _{ab}\mathcal{E}%
^{b}-2\delta ^{b}\mathcal{H}_{ab}\ ,
\end{equation}%
\begin{equation}
\widehat{\Sigma }_{\{ab\}}=\delta _{\{a}\Sigma _{b\}}+\frac{3\Sigma }{2}%
\zeta _{ab}-\varepsilon _{c\{a}\mathcal{H}_{b\}}^{\,\,\,\,\,c}\ ,
\end{equation}%
\begin{equation}
\widehat{\zeta }_{\{ab\}}=\left( \Sigma +\frac{\Theta }{3}\right) \Sigma
_{ab}+\delta _{\{a}a_{b\}}-\mathcal{E}_{ab}\ .
\end{equation}%
Finally, the constraints are%
\begin{equation}
\varepsilon ^{ab}\delta _{a}\mathcal{A}_{b}=0\ ,  \label{Abconstr}
\end{equation}%
\begin{equation}
p_{a}=-\left( \mu +p\right) \mathcal{A}_{a}\ ,
\end{equation}%
\begin{equation}
\varepsilon ^{ab}\delta _{a}\Sigma _{b}=-3\Sigma \xi +\mathcal{H}\ ,
\end{equation}%
\begin{equation}
\varepsilon _{ab}\delta ^{b}\textcolor{black}{\xi}+\!\!\delta ^{b}\zeta
_{ab}\!-\!\frac{\delta _{a}\phi }{2}=\frac{1}{2}\left( \!\Sigma \!-\frac{%
2\Theta }{3}\!\right) \!\Sigma _{a}\!+\mathcal{E}_{a},
\end{equation}%
\begin{equation}
\left( V_{a}-\frac{2}{3}W_{a}\right) +2\delta ^{b}\Sigma _{ab}=-2\varepsilon
_{ab}\mathcal{H}^{b}\ .  \label{constr3}
\end{equation}

The equations (\ref{phidot})-(\ref{constr3}) were derived from the generic
1+1+2 equations given in Ref. \cite{1+1+2}, by use of the commutation
relations given in Appendix \ref{commutation}. 

\subsection{Harmonic expansion\label{Harmonic}}

Following Ref. \cite{KASAscalar} we expand the scalar perturbation variables
into harmonics as 
\begin{equation}
\Psi =\displaystyle\sum\limits_{k_{\parallel },k_{\perp }}\Psi
_{k_{\parallel }k_{\perp }}^{S}\ P^{k_{\parallel }}\ Q^{k_{\perp }}\ .
\end{equation}%
The coefficients $\Psi _{k_{\parallel }k_{\perp }}^{S}$ depend solely of
time. The function $P^{k_{\parallel }}$ is the eigenfunction of the
Laplacian $\widehat{\Delta }=n^{a}\nabla _{a}n^{b}\nabla _{b}$ and it is
constant on the $z=const$ hypersurfaces: 
\begin{equation}
\widehat{\Delta }P^{k_{\parallel }}=-\frac{k_{\parallel }^{2}}{a_{1}^{2}}%
P^{k_{\parallel }}\ ,\ \delta _{a}P^{k_{\parallel }}=\dot{P}^{k_{\parallel
}}=0\ .
\end{equation}%
Here $k_{\parallel }$ are the constant comoving wave numbers in the
direction of anisotropy and the scale factor $a_{1}$ in this direction obeys%
\begin{equation}
\frac{\dot{a}_{1}}{a_{1}}=\Sigma +\frac{\Theta }{3}\ .
\end{equation}

The harmonics are introduced on the 2-sphere as%
\begin{equation}
\delta ^{2}Q^{l,m}=-\frac{l(l+1)}{a_{2}^{2}}Q^{l,m}~,~\widehat{Q}^{l,m}=\dot{%
Q}^{l,m}=0~,
\end{equation}%
where $\delta ^{2}=\delta _{a}\delta ^{a}$, and the second scale factor $%
a_{2}$ satisfies%
\begin{equation}
\frac{\dot{a}_{2}}{a_{2}}=-\frac{1}{2}\left( \Sigma -\frac{2\Theta }{3}%
\right) \ .
\end{equation}%
For a given $l$ value the index $m$ runs from $-l$ to $l$. Due to the
symmetries of the background spacetime the index $m$ never appear
explicitly, therefore we will use the following notation:%
\begin{equation}
\delta ^{2}Q^{k_{\perp }}=-\frac{k_{\perp }^{2}}{a_{2}^{2}}Q^{k_{\perp }}\
,\ \widehat{Q}^{k_{\perp }}=\dot{Q}^{k_{\perp }}=0\ ,
\end{equation}%
with $k_{\perp }^{2}=l(l+1)$ comoving wave numbers in the perpendicular
direction to $n^{a}$.

The vectors and tensors can be also expanded in harmonics by introducing the
vector and tensor spherical harmonics \cite{Schperturb,Schperturb2,LRSIItensor}.
The even (electric) and odd (magnetic) parity vector harmonics are%
\begin{equation}
Q_{a}^{k_{\perp }}=a_{2}\delta _{a}Q^{k_{\perp }}\ ,\ \ \overline{Q}%
_{a}^{k_{\perp }}=a_{2}\varepsilon _{ab}\delta ^{b}Q^{k_{\perp }}\ ,
\end{equation}%
and the vector $\Psi _{a}$ can be expanded as%
\begin{equation}
\Psi _{a}=\displaystyle\sum\limits_{k_{\parallel },k_{\perp
}}P^{k_{\parallel }}\ \left( \Psi _{k_{\parallel }k_{\perp
}}^{V}Q_{a}^{k_{\perp }}+\overline{\Psi }_{k_{\parallel }k_{\perp }}^{V}
\overline{Q}_{a}^{k_{\perp }}\right) \ .  \label{harmexpV}
\end{equation}%
Similarly, the even and odd tensor spherical harmonics are%
\begin{equation}
Q_{ab}^{k_{\perp }}=a_{2}^{2}\delta _{\{a}\delta _{b\}}Q^{k_{\perp }}\ ,\ 
\overline{Q}_{ab}^{k_{\perp }}=a_{2}^{2}\varepsilon _{c\{a}\delta ^{c}\delta
_{b\}}Q^{k_{\perp }}\ ,
\end{equation}%
and the tensor $\Psi _{ab}$ can be expanded as%
\begin{equation}
\Psi _{ab}=\displaystyle\sum\limits_{k_{\parallel },k_{\perp
}}P^{k_{\parallel }}\ \left( \Psi _{k_{\parallel }k_{\perp
}}^{T}Q_{ab}^{k_{\perp }}+\overline{\Psi }_{k_{\parallel }k_{\perp }}^{T}
\overline{Q}_{ab}^{k_{\perp }}\right) \ .
\end{equation}
{\color{black} This decomposition of the vectors and tensors encompasses both an
expansion into spherical harmonics and into even and odd modes, similarly to
the Regge-Wheeler decomposition of the perturbations of spherically
symmetric spacetimes, leading to the Regge-Wheeler equation for the odd
modes \cite{Regge} and the Zerilli equation for the even modes \cite{Zerilli,Zerilli2}. In our decomposition however the coefficients exhibit a
$(t,z)$ dependence, rather than $(t,r)$. Some useful relations involving the vector and tensor spherical harmonics
are enlisted in Appendix \ref{harmonics}.}

\subsection{Relations between harmonic coefficients from commutation rules
of covariant derivatives}

\label{harmonicrelations} 
A Kantowski-Sachs spacetime filled with perfect fluid is characterised by
the time-dependent scalars $G\equiv \left\{ \mu ,\mathcal{E},\Sigma ,\Theta
,p\right\} $. We expand their $\delta _{a}$-derivatives into harmonics on
the perturbed spacetime. Thus, applying the commutation relation Eq. (\ref%
{scalcomm4}) for zero-order scalars and using (\ref{vectspherid4}), we find
in the absence of vorticities the following relations:%
\begin{equation}
\overline{\mu }_{k_{\parallel }k_{\perp }}^{V}=\overline{X}_{k_{\parallel
}k_{\perp }}^{V}=\overline{V}_{k_{\parallel }k_{\perp }}^{V}=\overline{W}%
_{k_{\parallel }k_{\perp }}^{V}=\overline{p}_{k_{\parallel }k_{\perp
}}^{V}=0\ .  \label{OmegaSrel}
\end{equation}

However, we could also expand the anisotropic direction derivatives of $G$
into scalar harmonics, as they are also first-order. This expansion is%
\begin{equation}
\widehat{G}=\displaystyle\sum\limits_{k_{\parallel }k_{\perp }}\widetilde{G}%
_{k_{\parallel }k_{\perp }}^{S}\ P^{k_{\parallel }}\ Q^{k_{\perp }}\ .
\label{harmexp2}
\end{equation}%
Then the even parity part of Eq. (\ref{scalcomm3}) gives%
\begin{equation}
\frac{\widetilde{G}_{k_{\parallel }k_{\perp }}^{S}}{a_{2}}=\frac{%
ik_{\parallel }}{a_{1}}G_{k_{\parallel }k_{\perp }}^{V}\ ,
\label{harmexpsrel}
\end{equation}%
a constraint on $\widetilde{G}_{k_{\parallel }k_{\perp }}^{S}$ and $%
G_{k_{\parallel }k_{\perp }}^{V}$, emerging in the absence of vorticities.
Using Eq. (\ref{OmegaSrel}), the odd parity part of Eq. (\ref{scalcomm3})
becomes trivial. 

\subsection{Full set of evolution and constraint equations for the harmonic
coefficients\label{FullSet}}

From the commutation rules we have found that some harmonic coefficients
vanish in the absence of vorticities (see Eq. (\ref{OmegaSrel})). There is a
further coefficient $\overline{\mathcal{A}}_{k_{\parallel }k_{\perp }}^{V}=0$%
, the vanishing of which follows from Eq. (\ref{Abconstr}). Using these
relations, the perturbation equations (\ref{phidot})-(\ref{constr3}) can be
expanded into harmonics and are given in Appendix \ref{perturbEq_harmonics}.
Some of these equations are first integrals of the rest. We have found 17
independent constraints for 28 variables, as presented in Appendix \ref%
{perturbEq_harmonics}. We consider adiabatic matter perturbations $p=p\left(
\mu \right) $ giving $p_{k_{\parallel }k_{\perp }}^{V}=c_{s}^{2}\mu
_{k_{\parallel }k_{\perp }}^{V}$ with $c_{s}^{2}$ the square of the matter
speed of sound.\footnote{%
The adiabatic assumption for the total fluid perturbation, leading to $%
c_{s}^{2}=\dot{p}/\dot{\mu}$ is a good approximation when one of the matter
components dominates: for instance, $c_{s}^{2}=1/3$ in the radiation
dominated area and $c_{s}^{2}\approx 0$ for the dust dominated regime.}

In the frame associated to the fluid ($q_{a}=0$) the vorticity $\omega _{a}$
also vanishes because the fluid is irrotational. Moreover we have assumed $%
\pi _{ab}$ is negligible. These assumptions fixes completely the frame in a
1+3 covariant formalism, however do not in a 1+1+2 description where a dyad (%
$u^{a}$, $n^{a}$) must be assigned (see Appendix \ref{Frame}).\ The
quantities $q_{a}$, $\omega _{a}$ and $\pi _{ab}$ are invariant for some
part of the infinitesimal transformations which fixes $n^{a}$. In
particular, they are invariant under the infinitesimal translations given by 
$l_{a}$. We have 2 gauge degrees of freedom to sign $n^{a}$ perpendicularly
to $u^{a}$ on the perturbed spacetime and to fix completely the frame. In
the frame $a_{a}=0$ Eqs. (\ref{var7}) and (\ref{var8}) become constraints,
indicating that the fixing of $l_{a}$ reduces the degrees of freedom by
four. Thus, we have 6 degrees of freedom describing fully the vorticity-free
perturbations in the adiabatic case with $q_{a}=0=\pi _{ab}$. These
variables can be chosen as $\mu _{k_{\parallel }k_{\perp }}^{V}$, $\Sigma
_{k_{\parallel }k_{\perp }}^{T}$, $\mathcal{E}_{k_{\parallel }k_{\perp
}}^{T} $, $\overline{\mathcal{E}}_{k_{\parallel }k_{\perp }}^{T}$, $\mathcal{%
H}_{k_{\parallel }k_{\perp }}^{T}$ and $\overline{\mathcal{H}}_{k_{\parallel
}k_{\perp }}^{T}$. They are invariant for the $l_{a}$-infinitesimal
translation. Their evolutions are governed by two sets of decoupled
equations, which follow from the evolution equations (\ref{Eq1}), (\ref{Eq2}%
), (\ref{Eq3}), (\ref{Eq4}), (\ref{Eq5}) and (\ref{Eq6}) of Appendix \ref%
{perturbEq_harmonics} by employing the constraints. 

\subsubsection{Uncoupled evolutions of gravitational perturbations}\label{4.3.1}

The two coefficients $\overline{\mathcal{E}}_{k_{\parallel }k_{\perp }}^{T}$
and $\mathcal{H}_{k_{\parallel }k_{\perp }}^{T}$ form a decoupled system 
\begin{equation}
\dot{\overline{\mathcal{E}}}_{k_{\parallel }k_{\perp }}^{T}\!\!\!=-\frac{3}{2%
}\!\left( F\!+\!\Sigma D\!\right) \!\overline{\mathcal{E}}_{k_{\parallel
}k_{\perp }}^{T}\!\!\!+\frac{ik_{\parallel }}{a_{1}}\left( 1-D\right) 
\mathcal{H}_{k_{\parallel }k_{\perp }}^{T},  \label{4}
\end{equation}%
\begin{equation}
\dot{\mathcal{H}}_{k_{\parallel }k_{\perp }}^{T}=-\frac{a_{1}}{%
2ik_{\parallel }}\left( \frac{2k_{\parallel }^{2}}{a_{1}^{2}}-BC+9\Sigma
E\right) \overline{\mathcal{E}}_{k_{\parallel }k_{\perp }}^{T}-\frac{3}{2}%
\left( 2E+F\right) \mathcal{H}_{k_{\parallel }k_{\perp }}^{T}\ ,  \label{5}
\end{equation}%
with the coefficients%
\begin{equation}
B\equiv \frac{2k_{\parallel }^{2}}{a_{1}^{2}}+\frac{k_{\perp }^{2}}{a_{2}^{2}%
}+\frac{9\Sigma ^{2}}{2}+3\mathcal{E}=\frac{2k_{\parallel }^{2}}{a_{1}^{2}}-%
\frac{2-k_{\perp }^{2}}{a_{2}^{2}}+3\Sigma \left( \Sigma +\frac{\Theta }{3}%
\right) \ ,  \label{Bdef}
\end{equation}%
\begin{equation}
C\equiv B^{-1}\left( \frac{2-k_{\perp }^{2}}{a_{2}^{2}}+3\mathcal{E}\right)
\ ,  \label{Cdef}
\end{equation}%
\begin{equation}
D\equiv C+\frac{\mu +p}{B}\ ,
\end{equation}%
\begin{equation}
E\equiv \frac{\Sigma }{2}\left( C-\frac{\mathcal{E}}{B}\right) +\frac{\Theta 
\mathcal{E}}{3B}\ ,
\end{equation}%
\begin{equation}
F\equiv \Sigma +\frac{2\Theta }{3}\ .
\end{equation}%
The second equality in (\ref{Bdef}) follows from Eqs. (\ref{EWeyl}) and (\ref%
{2TimesGaussianCurv}).

Equivalently, the system can be rewritten as decoupled second-order linear
homogeneous ordinary differential equations:

\begin{equation}
\ddot{\overline{\mathcal{E}}}_{k_{\parallel }k_{\perp }}^{T}\!\!\!+q_{%
\overline{\mathcal{E}}1}\dot{\overline{\mathcal{E}}}_{k_{\parallel }k_{\perp
}}^{T}+q_{\overline{\mathcal{E}}0}\overline{\mathcal{E}}_{k_{\parallel
}k_{\perp }}^{T}=0~,  \label{waveE1}
\end{equation}%
\begin{equation}
\ddot{\mathcal{H}}_{k_{\parallel }k_{\perp }}^{T}+q_{\mathcal{H}1}\!\dot{%
\mathcal{H}}_{k_{\parallel }k_{\perp }}^{T}+q_{\mathcal{H}0}\mathcal{H}%
_{k_{\parallel }k_{\perp }}^{T}=0\ ,  \label{waveH1}
\end{equation}%
where%
\begin{equation}
q_{\overline{\mathcal{E}}1}=\frac{3}{2}\!\left( 2E+2F\!+\!\Sigma D\!\right) -%
\frac{d}{dt}\ln \frac{1-D}{a_{1}}~,
\end{equation}%
\begin{eqnarray}
2q_{\overline{\mathcal{E}}0} &=&\frac{1-D}{a_{1}}\left[ \frac{2k_{\parallel
}^{2}}{a_{1}}+a_{1}\left( 9\Sigma E-BC\right) \right] \!+3\frac{d}{dt}%
\!\left( F\!+\!\Sigma D\!\right)  \nonumber \\
&&-3\left( F\!+\!\Sigma D\!\right) \!\left[ \frac{d}{dt}\ln \frac{1-D}{a_{1}}%
-\frac{3}{2}\left( 2E+F\right) \right] ~,
\end{eqnarray}%
\begin{equation}
q_{\mathcal{H}1}=\frac{3}{2}\!\left( 2E+2F\!+\!\Sigma D\!\right) \!\!\!-%
\frac{d}{dt}\ln \left[ \frac{2k_{\parallel }^{2}}{a_{1}}+a_{1}\left( 9\Sigma
E-BC\right) \right] ~,
\end{equation}%
\begin{eqnarray}
2q_{\mathcal{H}0} &=&\frac{1-D}{a_{1}}\left[ \frac{2k_{\parallel }^{2}}{a_{1}%
}+a_{1}\left( 9\Sigma E-BC\right) \right] -3\left( 2E+F\right) \frac{d}{dt}%
\ln \left[ \frac{2k_{\parallel }^{2}}{a_{1}}+a_{1}\left( 9\Sigma E-BC\right) %
\right]  \nonumber \\
&&+\frac{9}{2}\!\left( F\!+\!\Sigma D\!\right) \left( 2E+F\right) +3\frac{d}{%
dt}\left( 2E+F\right) ~.
\end{eqnarray}

The equations (\ref{waveE1})-(\ref{waveH1}) represent wave equations with
friction. {\color{black} As will be shown in Section 5 of the paper, the quantities
obeying these equations represent the gravitational wave degrees of freedom.
}

\subsubsection{Evolutions with matter sources}\label{4.3.2}

The coefficients $\Sigma _{k_{\parallel }k_{\perp }}^{T}$, $\mathcal{E}
_{k_{\parallel }k_{\perp }}^{T}$ and $\overline{\mathcal{H}}_{k_{\parallel
}k_{\perp }}^{T}$ also form a system of differential equations coupled to
the density gradient $\mu _{k_{\parallel }k_{\perp }}^{V}$, as follows 
\begin{eqnarray}
\dot{\mu}_{k_{\parallel }k_{\perp }}^{V}\!\! &=&\left[ \frac{\Sigma }{2}
\left( 1-3\frac{\mu +p}{B}\right) \!-\frac{4\Theta }{3}\right] \mu
_{k_{\parallel }k_{\perp }}^{V}\!+\frac{a_{2}}{2}\left( \mu +p\right) 
\nonumber \\
&&\times \left[ \left( 1-C\right) \!\left( B\Sigma _{k_{\parallel }k_{\perp
}}^{T}\!-3\Sigma \mathcal{E}_{k_{\parallel }k_{\perp }}^{T}\right) -\frac{
a_{1}}{ik_{\parallel }}P\overline{\mathcal{H}}_{k_{\parallel }k_{\perp
}}^{T} \right] ~,  \label{eq11}
\end{eqnarray}
\begin{equation}
\dot{\Sigma}_{k_{\parallel }k_{\perp }}^{T}=-\frac{c_{s}^{2}}{a_{2}\left(
\mu +p\right) }\mu _{k_{\parallel }k_{\perp }}^{V}+\!\left( \Sigma -\frac{
2\Theta }{3}\right) \Sigma _{k_{\parallel }k_{\perp }}^{T}-\mathcal{E}
_{k_{\parallel }k_{\perp }}^{T}\ ,  \label{eq12}
\end{equation}
\begin{equation}
\dot{\mathcal{E}}_{k_{\parallel }k_{\perp }}^{T}\!=\!\frac{3\Sigma }{2a_{2}B}
\mu _{k_{\parallel }k_{\perp }}^{V}-\frac{\mu +p}{2}\Sigma _{k_{\parallel
}k_{\perp }}^{T}-\frac{3}{2}\left( F+\Sigma C\right) \mathcal{E}
_{k_{\parallel }k_{\perp }}^{T}+\frac{a_{1}}{2ik_{\parallel }}P\overline{ 
\mathcal{H}}_{k_{\parallel }k_{\perp }}^{T}\ ,  \label{eq13}
\end{equation}
\begin{equation}
\dot{\overline{\mathcal{H}}}_{k_{\parallel }k_{\perp }}^{T}\!\!\!\!=-\frac{
ik_{\parallel }}{a_{1}a_{2}B}\mu _{k_{\parallel }k_{\perp }}^{V}-\frac{ 
\overline{\mathcal{H}}_{k_{\parallel }k_{\perp }}^{T}}{S}-\frac{
ik_{\parallel }}{a_{1}}\left( 1-C\right) \mathcal{E}_{k_{\parallel }k_{\perp
}}^{T}\ .  \label{eq14}
\end{equation}
Here we have introduced the additional notations 
\begin{equation}
P\equiv \frac{2k_{\parallel }^{2}}{a_{1}^{2}}\left( 1-C\right) -\frac{%
k_{\perp }^{2}}{a_{2}^{2}}\frac{2-k_{\perp }^{2}}{a_{2}^{2}B}\ ,
\end{equation}%
\begin{equation}
S^{-1}\equiv \frac{2}{\Sigma B}\left[ \left( \mathcal{E}+\frac{3\Sigma ^{2}}{%
2}\right) \frac{2-k_{\perp }^{2}}{2a_{2}^{2}}-\mathcal{E}\frac{k_{\parallel
}^{2}}{a_{1}^{2}}\right] +\frac{3}{2}F+\frac{\mathcal{E}}{\Sigma }\ .
\end{equation}

This system is equivalent to the ones describing scalar perturbations
studied in \cite{KASAscalar}. Here the variables 
\begin{eqnarray}
{\mathcal{D}}_{a} &\equiv &a\frac{D_{a}\mu }{\mu }\,,\;\;{\mathcal{Z}}%
_{a}\equiv aD_{a}\Theta \,,\;\;{\mathcal{T}}_{a}\equiv aD_{a}\sigma ^{2}\,,
\\
\;\;{\mathcal{S}}_{a} &\equiv &D_{a}\left( \sigma ^{bc}S_{bc}\right)
\end{eqnarray}%
where $a$ is the average scale factor defined through $\Theta =3\dot{a}/a$
and $S_{ab}$ is the traceless part of the 3-Ricci tensor, were used. When
projected onto the 2-sphere and expressed in terms of the variables $\mu
_{a} $, $V_{a}$, $W_{a}$ and $X_{a}$ (see (\ref{gaugeinv})) they read 
\begin{eqnarray}
{\mathcal{D}}_{\bar{a}} &=&a\frac{\mu _{a}}{\mu }\,,\;\;{\mathcal{Z}}_{\bar{a%
}}=aW_{a}\,,\;\;{\mathcal{T}}_{\bar{a}}=\frac{3}{2}a\Sigma V_{a} \\
{\mathcal{S}}_{\bar{a}} &=&\frac{3a}{2}\Sigma X_{a}+a\left( \frac{3}{2}{%
\mathcal{E}}-\Theta \Sigma +\frac{9}{4}\Sigma ^{2}\right) V_{a}-\frac{a}{2}%
\Sigma ^{2}W_{a}\,.
\end{eqnarray}%
Due to equation (\ref{OmegaSrel}) we only have to consider the even parity
components, ${\mu }_{k_{\parallel }k_{\perp }}^{V}$, ${V}_{k_{\parallel
}k_{\perp }}^{V}$, ${W}_{k_{\parallel }k_{\perp }}^{V}$ and ${X}%
_{k_{\parallel }k_{\perp }}^{V}$, of the variables (\ref{gaugeinv}). The
three latter are solved for in terms of $\Sigma _{k_{\parallel }k_{\perp
}}^{T}$, $\mathcal{E}_{k_{\parallel }k_{\perp }}^{T}$, $\overline{\mathcal{H}%
}_{k_{\parallel }k_{\perp }}^{T}$ and $\mu _{k_{\parallel }k_{\perp }}^{V}$
in equations (\ref{VVconstr}), (\ref{WVconstr}) and (\ref{XVconstr}) in
appendix \ref{perturbEq_harmonics}. Substitution of these into equations
(C.1)-(C.4) in appendix C of \cite{KASAscalar} reproduces the system (\ref%
{eq11})-(\ref{eq14}).

We proceed with transforming Eqs. (\ref{eq12})-(\ref{eq14}) into second
order oscillator equations for each of the gravitational perturbations $%
\Sigma _{k_{\parallel }k_{\perp }}^{T}$, $\mathcal{E}_{k_{\parallel
}k_{\perp }}^{T}$ and $\overline{\mathcal{H}}_{k_{\parallel }k_{\perp }}^{T}$%
, with source terms given by the matter perturbations $\mu _{k_{\parallel
}k_{\perp }}^{V}$ and $\dot{\mu}_{k_{\parallel }k_{\perp }}^{V}$which induce
forced oscillations. In turn, then these gravitational perturbations act as
sources for the first order evolutions of $\mu _{k_{\parallel }k_{\perp
}}^{V}$. For the gravitational perturbations we obtain%
\begin{equation}
\ddot{\Sigma}_{k_{\parallel }k_{\perp }}^{T}+q_{\Sigma 1}\dot{\Sigma}%
_{k_{\parallel }k_{\perp }}^{T}+q_{\Sigma 0}\Sigma _{k_{\parallel }k_{\perp
}}^{T}=\frac{\left( 1-c_{s}^{2}\right) }{a_{2}\left( \mu +p\right) }\dot{\mu}%
_{k_{\parallel }k_{\perp }}^{V}+\frac{s_{\Sigma 0}}{a_{2}\left( \mu
+p\right) }\mu _{k_{\parallel }k_{\perp }}^{V}~,  \label{SigmaEq}
\end{equation}%
\begin{equation}
\ddot{\mathcal{E}}_{k_{\parallel }k_{\perp }}^{T}\!+q_{\mathcal{E}1}\dot{%
\mathcal{E}}_{k_{\parallel }k_{\perp }}^{T}+q_{\mathcal{E}0}\mathcal{E}%
_{k_{\parallel }k_{\perp }}^{T}=\frac{s_{\mathcal{E}1}}{a_{2}}\dot{\mu}%
_{k_{\parallel }k_{\perp }}^{V}+s_{\mathcal{E}0}\mu _{k_{\parallel }k_{\perp
}}^{V}\ ,  \label{calEEq}
\end{equation}%
and 
\begin{equation}
\ddot{\overline{\mathcal{H}}}_{k_{\parallel }k_{\perp }}^{T}\!\!\!\!+q_{%
\overline{\mathcal{H}}1}\dot{\overline{\mathcal{H}}}_{k_{\parallel }k_{\perp
}}^{T}+q_{\overline{\mathcal{H}}0}\overline{\mathcal{H}}_{k_{\parallel
}k_{\perp }}^{T}=\frac{ik_{\parallel }s_{\overline{\mathcal{H}}0}}{%
a_{1}a_{2}B}\mu _{k_{\parallel }k_{\perp }}^{V}\!\!\!\!\ ,  \label{calHbarEq}
\end{equation}%
with the coefficients%
\begin{equation}
q_{\Sigma 1}=2\Sigma +\frac{5\Theta }{3}~,
\end{equation}%
\begin{equation}
q_{\Sigma 0}=\frac{\left( 1-C\right) B-\left( \mu +p\right) }{2}\!-\frac{d}{%
dt}\!\left( \Sigma -\frac{2\Theta }{3}\right) -3\left( \Sigma +\frac{\Theta 
}{3}\right) \!\left( \Sigma -\frac{2\Theta }{3}\right) ~,
\label{SigmaCoeffn1}
\end{equation}%
\begin{equation}
s_{\Sigma 0}=\frac{4\Theta }{3}-\frac{\Sigma }{2}\!-c_{s}^{2}\left( \Theta
+3\Sigma \right) -\dot{c}_{s}^{2}+\frac{d}{dt}\ln \left[ a_{2}\left( \mu
+p\right) \right] ~,  \label{SigmaCoeffn2}
\end{equation}%
\begin{eqnarray}
q_{\mathcal{E}1} &=&W_{1}\left[ \frac{1}{S}\!\!+\!\Theta +\frac{3}{2}\Sigma
\left( 1+C\right) -\frac{d}{dt}\ln \left( a_{1}P\right) \right]  \nonumber \\
&&-W_{2}\left[ \frac{5\Theta }{3}+\frac{\Sigma }{2}\left( 1+3C\right) -\frac{%
d}{dt}\ln \left( \mu +p\right) \right] ~,  \label{calECoeffn1}
\end{eqnarray}%
\begin{eqnarray}
q_{\mathcal{E}0} &=&\frac{3}{2}\frac{d}{dt}\left( \Sigma +\frac{2\Theta }{3}%
+\Sigma C\right) +\frac{\left( 1-C\right) P}{2}-\frac{\mu +p}{2}%
-W_{1}\!\left( \Theta +\frac{3}{2}\Sigma \left( 1+C\right) \right)  \nonumber
\\
&&\times \left[ \frac{d}{dt}\ln \left( a_{1}P\right) -\frac{1}{S}\right]
+W_{2}\left( \!\Theta +3\Sigma \right) \left[ \Sigma -\frac{2\Theta }{3}+%
\frac{d}{dt}\ln \left( \mu +p\right) \right]  \nonumber \\
&&-W_{2}\frac{3\Sigma \left( 1-C\right) }{2}\left[ \frac{d}{dt}\ln \left(
a_{1}P\right) -\frac{1}{S}\right] \!~,
\end{eqnarray}%
\begin{eqnarray}
s_{\mathcal{E}1} &=&\frac{3\Sigma }{2B}\!\!+\frac{1}{\left( 1-C\right)
\!B\!-\left( \mu +p\right) }\left( \frac{2\Theta }{3}-\Sigma -\frac{1}{S}+%
\frac{d}{dt}\ln \frac{a_{1}P}{\mu +p}\right) ~, \\
s_{\mathcal{E}0} &=&\frac{d}{dt}\!\left( \frac{3\Sigma }{2a_{2}B}\right) -%
\frac{P}{2a_{2}B}+\frac{c_{s}^{2}}{2a_{2}}+\frac{W_{3}-W_{4}}{a_{2}\left[
\left( 1-C\right) \!B\!-\left( \mu +p\right) \right] }~,  \label{calECoeffn4}
\end{eqnarray}%
and%
\begin{eqnarray}
q_{\overline{\mathcal{H}}1} &=&\frac{1}{S}+Q~,  \label{calHbarCoeffn1} \\
q_{\overline{\mathcal{H}}0} &=&\frac{Q}{S}+\left( 1-C-\frac{\mu +p}{B}%
\right) \frac{P}{2}+\frac{d}{dt}\frac{1}{S}~, \\
s_{\overline{\mathcal{H}}0} &=&\frac{\Theta }{3}-\frac{7\Sigma }{2}\!+\frac{d%
}{dt}\ln \left[ a_{2}B\left( 1-C\right) \right] ~,  \label{calHbarCoeffn3}
\end{eqnarray}%
where we have denoted%
\begin{eqnarray}
W_{1} &=&\frac{\left( 1-C\right) \!B}{\left( 1-C\right) \!B\!-\left( \mu
+p\right) }~,  \label{W1} \\
W_{2} &=&\frac{\left( \mu +p\right) }{\left( 1-C\right) \!B\!-\left( \mu
+p\right) }~,  \label{W2}
\end{eqnarray}%
\begin{equation}
W_{3}=\left[ \frac{d}{dt}\ln \left( a_{1}P\right) -\frac{1}{S}\right] \left[ 
\frac{4\Theta }{3}-\!\frac{\Sigma \left( 4-3C\right) \!}{2}+\!\frac{3\Sigma
\left( \mu +p\right) }{2B}\right] ~,  \label{W3}
\end{equation}%
\begin{eqnarray}
W_{4} &=&\left[ \frac{d}{dt}\ln \left( \mu +p\right) +\left( \Sigma -\frac{%
2\Theta }{3}\right) \right] \left( \frac{4\Theta }{3}-\frac{\Sigma }{2}%
\right) ~,  \label{W4} \\
Q &=&\Theta +\frac{3\Sigma }{2}\left( 1+C+\frac{\mu +p}{B}\right) -\frac{d}{%
dt}\ln \frac{1-C}{a_{1}}~.  \label{Q}
\end{eqnarray}

In the next section we analyse the high frequency limit of these equations. 

\section{Geometrical optics approximation\label{GeomOp}}

In this section we follow Isaacson's definition \cite{Isaacson1}, \cite%
{Isaacson2} of gravitational waves on a curved background in a geometrical
optics approximation. The key concept is that gravitational waves are
periodic perturbations with a wavelength much shorter than the curvature
radius of the background. This is known as the geometrical optics
approximation, or the high frequency limit. In the notations of the present
paper the physical wave numbers along $z$ and along the spheres are $%
k_{\parallel }/a_{1}$ and $k_{\perp }/a_{2}$, respectively.

Then $k_{\parallel }$, $k_{\perp }\gg 1$, while a glance on Eqs. (\ref%
{Thetabackgroundexp}), (\ref{Sigmabackgroundexp}), (\ref{EWeyl}) and (\ref%
{2TimesGaussianCurv}) implies 
\begin{equation}
L\left( \frac{2k_{\parallel }^{2}}{a_{1}^{2}},\frac{k_{\perp }^{2}}{%
a_{2}^{2} }\right) \gg \Theta ^{2},\Sigma ^{2},\mathcal{E},\mu ,p~,
\label{kpera}
\end{equation}
where $L$ is any linear combination with coefficients of order unity.
Implementing these in the equations we get 
\begin{equation}
B\simeq P\simeq \frac{2k_{\parallel }^{2}}{a_{1}^{2}}+\frac{k_{\perp }^{2}}{
a_{2}^{2}}\ ,  \label{Bgeom}
\end{equation}
and 
\begin{equation}
D\simeq C\simeq -B^{-1}\frac{k_{\perp }^{2}}{a_{2}^{2}}\ ,\quad E\simeq 
\frac{\Sigma }{2}C\ ,  \label{CDEgeom}
\end{equation}
\begin{equation}
S\simeq -\Sigma B\left[ \left( \mathcal{E}+\frac{3\Sigma ^{2}}{2}\right) 
\frac{k_{\perp }^{2}}{a_{2}^{2}}+2\mathcal{E}\frac{k_{\parallel }^{2}}{
a_{1}^{2}}\right] ^{-1}\ .
\end{equation}
Thus in the geometrical optics approximation the order of the dimensionless
quantities relates as $\mathcal{O}\left( a_{i}^{2}B\right) =\mathcal{O}
\left( a_{i}^{2}P\right) \gg \mathcal{O}\left( C\right) =\mathcal{O}\left(
D\right) =\mathcal{O}\left( a_{i}E\right) =\mathcal{O}\left( a_{i}F\right) = 
\mathcal{O}\left( a_{i}^{-1}S\right) =\mathcal{O}\left( 1\right) $. 

\subsection{High frequency evolutions of the uncoupled gravitational
perturbations: gravitational waves}

The relevant coefficients are approximated as 
\begin{eqnarray}
q_{\overline{\mathcal{E}}0} &\simeq &\!q_{\mathcal{H}0}\simeq \!\left(
1-C\right) \!\!\left( \frac{k_{\parallel }^{2}}{a_{1}^{2}}+\frac{k_{\perp
}^{2}}{2a_{2}^{2}}\right) =\frac{k_{\parallel }^{2}}{a_{1}^{2}}+\frac{%
k_{\perp }^{2}}{a_{2}^{2}}, \\
q_{\overline{\mathcal{E}}1} &\simeq &3\left( F\!+\!\Sigma C\!\right) -\frac{d%
}{dt}\ln \frac{1-C}{a_{1}}~, \\
q_{\mathcal{H}1} &\simeq &3\left( F\!+\!\Sigma C\!\right) -\frac{d}{dt}\ln %
\left[ a_{1}\left( \frac{2k_{\parallel }^{2}}{a_{1}^{2}}+\frac{k_{\perp }^{2}%
}{a_{2}^{2}}\right) \right] .
\end{eqnarray}%
The damped wave equations (\ref{waveE1})-(\ref{waveH1}) simplify to%
\begin{equation}
\ddot{\overline{\mathcal{E}}}_{k_{\parallel }k_{\perp }}^{T}\!\!\!+q_{%
\overline{\mathcal{E}}1}\dot{\overline{\mathcal{E}}}_{k_{\parallel }k_{\perp
}}^{T}+\left( \frac{k_{\parallel }^{2}}{a_{1}^{2}}+\frac{k_{\perp }^{2}}{%
a_{2}^{2}}\right) \overline{\mathcal{E}}_{k_{\parallel }k_{\perp }}^{T}=0~,
\end{equation}%
\begin{equation}
\ddot{\mathcal{H}}_{k_{\parallel }k_{\perp }}^{T}+q_{\mathcal{H}1}\!\dot{%
\mathcal{H}}_{k_{\parallel }k_{\perp }}^{T}+\left( \frac{k_{\parallel }^{2}}{%
a_{1}^{2}}+\frac{k_{\perp }^{2}}{a_{2}^{2}}\right) \mathcal{H}_{k_{\parallel
}k_{\perp }}^{T}=0\ .
\end{equation}%
Both of these equations are of the form $\ddot{X}+2\zeta \Omega \dot{X}%
+\Omega ^{2}X=0$, where $\Omega $ represents the undamped angular frequency.
For $\zeta <1$ the oscillator is underdamped and the real angular frequency
is given by $\Omega \sqrt{1-\zeta ^{2}}$. The propagation speed of the wave
therefore is $c_{w}=\Omega \sqrt{1-\zeta ^{2}}/k_{phys}$ (with $k_{phys}$
given by $k_{\parallel }/a_{1}$ or $k_{\perp }/a_{2}$, when the propagation
is along $z$ or along the spheres). These considerations imply that we are
assuming a negligible change in the scale factors over one period.

In order to continue the analysis we define a small parameter $\varepsilon
\approx \left( a_{i}k_{phys}\right) ^{-1}$ (characterising the geometrical
optics approximation) and we consider perturbations along the $z$ direction
and along the sphere separately. 

\subsubsection{Waves propagating along the $z$-direction}

We get $C\simeq 0\ $and $F\!+\!\Sigma C\!=\frac{2\Theta }{3}+\Sigma ,$ hence
the propagation equations are%
\begin{equation}
\ddot{\overline{\mathcal{E}}}_{k_{\parallel }k_{\perp }}^{T}\!\!\!+\left(
2\Theta +3\Sigma +\frac{\dot{a}_{1}}{a_{1}}\right) \dot{\overline{\mathcal{E}%
}}_{k_{\parallel }k_{\perp }}^{T}+\frac{k_{\parallel }^{2}}{a_{1}^{2}}%
\overline{\mathcal{E}}_{k_{\parallel }k_{\perp }}^{T}=0~,
\end{equation}%
\begin{equation}
\ddot{\mathcal{H}}_{k_{\parallel }k_{\perp }}^{T}+\left( 2\Theta +3\Sigma +%
\frac{\dot{a}_{1}}{a_{1}}\right) \!\dot{\mathcal{H}}_{k_{\parallel }k_{\perp
}}^{T}+\frac{k_{\parallel }^{2}}{a_{1}^{2}}\mathcal{H}_{k_{\parallel
}k_{\perp }}^{T}=0\ .
\end{equation}%
The damping parameter turns out to be 
\begin{equation}
\zeta _{\parallel }=\frac{a_{1}}{2k_{\parallel }}\left( 2\Theta +3\Sigma +%
\frac{\dot{a}_{1}}{a_{1}}\right) =\mathcal{O}\left( \varepsilon \right) ~,
\end{equation}%
and the speed of propagation of both $\overline{\mathcal{E}}_{k_{\parallel
}k_{\perp }}^{T}$ and $\mathcal{H}_{k_{\parallel }k_{\perp }}^{T}$ is%
\begin{equation}
c_{\parallel }=\sqrt{1-\zeta ^{2}}\simeq 1-\frac{\zeta ^{2}}{2}=1-\mathcal{O}%
\left( \varepsilon ^{2}\right) ~.
\end{equation}%
Thus to linear order in the geometrical optics approximation both $\overline{%
\mathcal{E}}_{k_{\parallel }k_{\perp }}^{T}$ and $\mathcal{H}_{k_{\parallel
}k_{\perp }}^{T}$ represent gravitational waves propagating with the speed
of light. 

\subsubsection{Waves propagating along the spheres}

We get $C\simeq -1\ $and $F\!+\!\Sigma C\!=\frac{2\Theta }{3}$, hence 
\begin{equation}
\ddot{\overline{\mathcal{E}}}_{k_{\parallel }k_{\perp }}^{T}\!\!\!+\left(
2\Theta +\frac{\dot{a}_{1}}{a_{1}}\right) \dot{\overline{\mathcal{E}}}%
_{k_{\parallel }k_{\perp }}^{T}+\frac{k_{\perp }^{2}}{a_{2}^{2}}\overline{%
\mathcal{E}}_{k_{\parallel }k_{\perp }}^{T}=0~,
\end{equation}%
\begin{equation}
\ddot{\mathcal{H}}_{k_{\parallel }k_{\perp }}^{T}+\left( 2\Theta -\frac{\dot{%
a}_{1}}{a_{1}}+\frac{2\dot{a}_{2}}{a_{2}}\right) \!\dot{\mathcal{H}}%
_{k_{\parallel }k_{\perp }}^{T}+\frac{k_{\perp }^{2}}{a_{2}^{2}}\mathcal{H}%
_{k_{\parallel }k_{\perp }}^{T}=0\ .
\end{equation}%
Then there are different damping parameters for the two fields: 
\begin{eqnarray}
\zeta _{\perp \overline{\mathcal{E}}} &=&\frac{a_{2}}{2k_{\perp }}\left(
2\Theta +\frac{\dot{a}_{1}}{a_{1}}\right) =\mathcal{O}\left( \varepsilon
\right) ~, \\
\zeta _{\perp \mathcal{H}} &=&\frac{a_{2}}{2k_{\perp }}\left( 2\Theta -\frac{%
\dot{a}_{1}}{a_{1}}+\frac{2\dot{a}_{2}}{a_{2}}\right) =\mathcal{O}\left(
\varepsilon \right) ~,
\end{eqnarray}%
and the speeds of propagation of $\overline{\mathcal{E}}_{k_{\parallel
}k_{\perp }}^{T}$ and $\mathcal{H}_{k_{\parallel }k_{\perp }}^{T}$ are also
different, but only to second order:%
\begin{eqnarray}
c_{\perp \overline{\mathcal{E}}} &\simeq &1-\frac{\zeta _{\overline{\mathcal{%
E}}}^{2}}{2}=1-\mathcal{O}\left( \varepsilon ^{2}\right) ~, \\
c_{\perp \mathcal{H}} &\simeq &1-\frac{\zeta _{\mathcal{H}}^{2}}{2}=1-%
\mathcal{O}\left( \varepsilon ^{2}\right) ~.
\end{eqnarray}%
Again, to linear order in the geometrical optics approximation both $%
\overline{\mathcal{E}}_{k_{\parallel }k_{\perp }}^{T}$ and $\mathcal{H}%
_{k_{\parallel }k_{\perp }}^{T}$ represent gravitational waves propagating
with the speed of light. This is consistent with the generic theory, where
the gravitational waves appear at the first order of the expansion of the
Einstein equations. The second order terms, neglected in this picture can be
interpreted as backreaction, leading to wavenumber-dependent dispersion. 

\subsection{High frequency evolutions with matter sources: gravitational,
shear and matter waves}

To first order in the geometrical optics approximation, the shorthand
notations appearing in the evolution equations (\ref{SigmaEq})-(\ref%
{calHbarEq}) simplify as follows. The coefficients of the algebraic terms of
the gravitational perturbations are%
\begin{equation}
q_{\Sigma 0}\simeq q_{\mathcal{E}0}\simeq q_{\overline{\mathcal{H}}0}\simeq 
\frac{k_{\parallel }^{2}}{a_{1}^{2}}+\frac{k_{\perp }^{2}}{a_{2}^{2}}\!~,
\end{equation}%
the coefficients of the damping terms become%
\begin{equation}
q_{\Sigma 1}\simeq 2\Sigma +\frac{5\Theta }{3}~,
\end{equation}%
\begin{equation}
q_{\mathcal{E}1}=\Theta +\frac{\left( \mathcal{E}+\frac{3\Sigma ^{2}}{2}%
\right) \frac{k_{\perp }^{2}}{a_{2}^{2}}+\left( 2\mathcal{E}+3\Sigma
^{2}\right) \frac{k_{\parallel }^{2}}{a_{1}^{2}}}{\Sigma \left( \frac{%
2k_{\parallel }^{2}}{a_{1}^{2}}+\frac{k_{\perp }^{2}}{a_{2}^{2}}\right) }-%
\frac{d}{dt}\ln \left[ a_{1}\left( \frac{2k_{\parallel }^{2}}{a_{1}^{2}}+%
\frac{k_{\perp }^{2}}{a_{2}^{2}}\right) \right] ~,
\end{equation}%
\begin{equation}
q_{\overline{\mathcal{H}}1}=\Theta +\frac{\left( \mathcal{E}+\frac{3\Sigma
^{2}}{2}\right) \frac{k_{\perp }^{2}}{a_{2}^{2}}+\left( 2\mathcal{E}+3\Sigma
^{2}\right) \frac{k_{\parallel }^{2}}{a_{1}^{2}}}{\Sigma \left( \frac{%
2k_{\parallel }^{2}}{a_{1}^{2}}+\frac{k_{\perp }^{2}}{a_{2}^{2}}\right) }-%
\frac{d}{dt}\ln \frac{\frac{k_{\parallel }^{2}}{a_{1}^{2}}+\frac{k_{\perp
}^{2}}{a_{2}^{2}}}{a_{1}\left( \frac{2k_{\parallel }^{2}}{a_{1}^{2}}+\frac{%
k_{\perp }^{2}}{a_{2}^{2}}\right) }~,
\end{equation}%
the coefficients in the algebraic source terms read%
\begin{equation}
s_{\Sigma 0}=\frac{4\Theta }{3}-\frac{\Sigma }{2}\!-c_{s}^{2}\left( \Theta
+3\Sigma \right) -\dot{c}_{s}^{2}+\frac{d}{dt}\ln \left[ a_{2}\left( \mu
+p\right) \right] ~,
\end{equation}%
\begin{equation}
s_{\mathcal{E}0}=-\frac{1-c_{s}^{2}}{2a_{2}}~,
\end{equation}%
\begin{equation}
s_{\overline{\mathcal{H}}0}=\frac{\Theta }{3}-\frac{7\Sigma }{2}\!+\frac{d}{%
dt}\ln \left[ a_{2}\left( \frac{k_{\parallel }^{2}}{a_{1}^{2}}+\frac{%
k_{\perp }^{2}}{a_{2}^{2}}\right) \right] ~,
\end{equation}%
while the coefficient of the time derivative source term in Eq. (\ref{calEEq}%
) is%
\begin{eqnarray}
s_{\mathcal{E}1} &=&\frac{1}{2}\left( \frac{k_{\parallel }^{2}}{a_{1}^{2}}+%
\frac{k_{\perp }^{2}}{a_{2}^{2}}\right) ^{-1}\left[ \frac{d}{dt}\ln \frac{%
a_{1}\left( \frac{2k_{\parallel }^{2}}{a_{1}^{2}}+\frac{k_{\perp }^{2}}{%
a_{2}^{2}}\right) }{\mu +p}\right.  \nonumber \\
&&\left. +\frac{\left( \frac{7\Sigma ^{2}}{2}+\frac{2\Theta \Sigma }{3}+%
\mathcal{E}\right) \frac{k_{\perp }^{2}}{a_{2}^{2}}+\left( 2\Sigma ^{2}+%
\frac{2\Theta \Sigma }{3}+\mathcal{E}\right) \frac{2k_{\parallel }^{2}}{%
a_{1}^{2}}}{\Sigma \left( \frac{2k_{\parallel }^{2}}{a_{1}^{2}}+\frac{%
k_{\perp }^{2}}{a_{2}^{2}}\right) }\right] ~.
\end{eqnarray}%
Hence the evolutions (\ref{SigmaEq})-(\ref{calHbarEq}), to leading order
simplify as%
\begin{equation}
\ddot{\Sigma}_{k_{\parallel }k_{\perp }}^{T}+q_{\Sigma 1}\dot{\Sigma}%
_{k_{\parallel }k_{\perp }}^{T}+\left( \frac{k_{\parallel }^{2}}{a_{1}^{2}}+%
\frac{k_{\perp }^{2}}{a_{2}^{2}}\right) \Sigma _{k_{\parallel }k_{\perp
}}^{T}=\frac{\left( 1-c_{s}^{2}\right) }{a_{2}\left( \mu +p\right) }\dot{\mu}%
_{k_{\parallel }k_{\perp }}^{V}~,  \label{SigmaW1}
\end{equation}%
\begin{equation}
\ddot{\mathcal{E}}_{k_{\parallel }k_{\perp }}^{T}\!+q_{\mathcal{E}1}\dot{%
\mathcal{E}}_{k_{\parallel }k_{\perp }}^{T}+\left( \frac{k_{\parallel }^{2}}{%
a_{1}^{2}}+\frac{k_{\perp }^{2}}{a_{2}^{2}}\right) \mathcal{E}_{k_{\parallel
}k_{\perp }}^{T}=0\ ,  \label{calEW}
\end{equation}%
\begin{equation}
\ddot{\overline{\mathcal{H}}}_{k_{\parallel }k_{\perp }}^{T}\!\!\!\!+q_{%
\overline{\mathcal{H}}1}\dot{\overline{\mathcal{H}}}_{k_{\parallel }k_{\perp
}}^{T}+\left( \frac{k_{\parallel }^{2}}{a_{1}^{2}}+\frac{k_{\perp }^{2}}{%
a_{2}^{2}}\right) \overline{\mathcal{H}}_{k_{\parallel }k_{\perp
}}^{T}=0\!\!\!\!\ ,  \label{calHbarW}
\end{equation}%
The fourth equation of the closed system becomes%
\begin{equation}
\dot{\mu}_{k_{\parallel }k_{\perp }}^{V}\!\!=a_{2}\left( \mu +p\right) \left[
\left( \frac{k_{\parallel }^{2}}{a_{1}^{2}}+\frac{k_{\perp }^{2}}{a_{2}^{2}}%
\right) \Sigma _{k_{\parallel }k_{\perp }}^{T}\!-\frac{a_{1}}{2ik_{\parallel
}}\left( \frac{2k_{\parallel }^{2}}{a_{1}^{2}}+\frac{k_{\perp }^{2}}{%
a_{2}^{2}}\right) \overline{\mathcal{H}}_{k_{\parallel }k_{\perp }}^{T}%
\right] ~.  \label{mudot}
\end{equation}%
Inserting Eq. (\ref{mudot}) into Eq. (\ref{SigmaW1}) we obtain%
\begin{equation}
\ddot{\Sigma}_{k_{\parallel }k_{\perp }}^{T}+q_{\Sigma 1}\dot{\Sigma}%
_{k_{\parallel }k_{\perp }}^{T}+c_{s}^{2}\left( \frac{k_{\parallel }^{2}}{%
a_{1}^{2}}+\frac{k_{\perp }^{2}}{a_{2}^{2}}\right) \Sigma _{k_{\parallel
}k_{\perp }}^{T}=-\frac{a_{1}\left( 1-c_{s}^{2}\right) }{2ik_{\parallel }}%
\left( \frac{2k_{\parallel }^{2}}{a_{1}^{2}}+\frac{k_{\perp }^{2}}{a_{2}^{2}}%
\right) \overline{\mathcal{H}}_{k_{\parallel }k_{\perp }}^{T}~,
\label{SigmaW}
\end{equation}%
We comment on the system (\ref{calEW})-(\ref{SigmaW}) as follows. {%
\color{black} The most striking feature is that the gravitational sector $%
\mathcal{E}_{k_{\parallel }k_{\perp }}^{T}$ and $\overline{\mathcal{H}}%
_{k_{\parallel }k_{\perp }}^{T}$ fully decouples from the matter density
gradient $\mu _{k_{\parallel }k_{\perp }}^{V}$.} While $\mathcal{E}%
_{k_{\parallel }k_{\perp }}^{T}$ and $\overline{\mathcal{H}}_{k_{\parallel
}k_{\perp }}^{T}$ obey damped oscillator equations (similarly to their
counterparts $\overline{\mathcal{E}}_{k_{\parallel }k_{\perp }}^{T}$ and $%
\mathcal{H}_{k_{\parallel }k_{\perp }}^{T}$, discussed in the previous
subsection), $\Sigma _{k_{\parallel }k_{\perp }}^{T}$ undergoes a forced
oscillation. Let us first discuss the damped oscillations in the manner of
the previous subsection. 

\subsubsection{Gravitational waves}

The undamped angular frequencies of $\overline{\mathcal{E}}_{k_{\parallel
}k_{\perp }}^{T}$ and $\mathcal{H}_{k_{\parallel }k_{\perp }}^{T}$ are $%
\Omega =\left( \frac{k_{\parallel }^{2}}{a_{1}^{2}}+\frac{k_{\perp }^{2}}{%
a_{2}^{2}}\right) ^{1/2}$, while the damping factors are $\zeta _{\mathcal{E}%
}=\frac{q_{\mathcal{E}1}}{2\Omega }$ and $\zeta _{\overline{\mathcal{H}}}=%
\frac{q_{\overline{\mathcal{H}}1}}{2\Omega }$. For the waves propagating in
the $z$ and spherical directions, respectively, we get the following damping
factors of $\mathcal{O}\left( \varepsilon \right) $:%
\begin{equation}
\zeta _{\parallel \mathcal{E}}=\zeta _{\parallel \overline{\mathcal{H}}}=%
\frac{a_{1}}{2k_{\parallel }}\left( \Theta +\frac{2\mathcal{E}+3\Sigma ^{2}}{%
2\Sigma }+\frac{d}{dt}\ln a_{1}\right)
\end{equation}%
and%
\begin{eqnarray}
\zeta _{\perp \mathcal{E}} &=&\frac{a_{2}}{2k_{\perp }}\left( \Theta +\frac{2%
\mathcal{E}+3\Sigma ^{2}}{2\Sigma }-\frac{d}{dt}\ln \frac{a_{1}}{a_{2}^{2}}%
\right) ~, \\
\zeta _{\perp \overline{\mathcal{H}}} &=&\frac{a_{2}}{2k_{\perp }}\left(
\Theta +\frac{2\mathcal{E}+3\Sigma ^{2}}{2\Sigma }+\frac{d}{dt}\ln
a_{1}\right) ~,
\end{eqnarray}%
also the corresponding propagation speeds:%
\begin{eqnarray}
c_{\parallel \mathcal{E}} &\simeq &1-\frac{\zeta _{\parallel \mathcal{E}}^{2}%
}{2}=1-\mathcal{O}\left( \varepsilon ^{2}\right) ~, \\
c_{\parallel \overline{\mathcal{H}}} &\simeq &1-\frac{\zeta _{\parallel 
\overline{\mathcal{H}}}^{2}}{2}=1-\mathcal{O}\left( \varepsilon ^{2}\right)
\end{eqnarray}%
and 
\begin{eqnarray}
c_{\perp \mathcal{E}} &\simeq &1-\frac{\zeta _{\perp \mathcal{E}}^{2}}{2}=1-%
\mathcal{O}\left( \varepsilon ^{2}\right) ~, \\
c_{\perp \overline{\mathcal{H}}} &\simeq &1-\frac{\zeta _{\perp \overline{%
\mathcal{H}}}^{2}}{2}=1-\mathcal{O}\left( \varepsilon ^{2}\right) ~.
\end{eqnarray}%
Thus, to leading order in the geometrical optics approximation the
gravito-magnetic variables $\overline{\mathcal{E}}_{k_{\parallel }k_{\perp
}}^{T}$ and $\mathcal{H}_{k_{\parallel }k_{\perp }}^{T}$ represent pure
gravitational waves. The non-identical corrections at higher order represent
backreaction. 

\subsubsection{Shear waves and matter density gradient waves}

To leading order in the geometrical optics approximation Eq. (\ref{SigmaW})
represents a wave for the shear $\Sigma _{k_{\parallel }k_{\perp }}^{T}$
propagating with the speed of sound $c_{s}$. At higher order both a damping
mechanism and a force acts on this wave. The force is generated by the
gravitational wave degree of freedom $\overline{\mathcal{H}}_{k_{\parallel
}k_{\perp }}^{T}$.

Again to leading order the shear $\Sigma _{k_{\parallel }k_{\perp }}^{T}$ is
but a time derivative of the matter density gradient $\mu _{k_{\parallel
}k_{\perp }}^{V}$. {\color{black} In order to understand this claim it is
necessary to remember that in the geometrical optics limit we consider
wavelengths much shorter than the curvature radius, hence for the purpose of
the wave propagation we can approximate the background scale factors and
fluid characteristics as constants. }Hence the latter also represents a wave
propagating with the speed of sound $c_{s}$, but dephased with an angle $\pi
/2$. This is in agreement with our assumption of an adiabatic speed of sound
given by $p_{k_{\parallel }k_{\perp }}^{V}=c_{s}^{2}\mu _{k_{\parallel
}k_{\perp }}^{V}$.

To higher order it mimics the damped and forced oscillation of $\Sigma
_{k_{\parallel }k_{\perp }}^{T}$. 

\subsection{{\color{black} The degrees of freedom in the gravitational waves}}

{\color{black} In the geometrical optics approximation we have obtained four equations
representing gravitational waves propagating with the speed of light, for
the even and odd modes of the 2D electric and magnetic projections of the
Weyl tensor, $\mathcal{E}_{k_{\parallel }k_{\perp }}^{T}$, $\overline{
\mathcal{E}}_{k_{\parallel }k_{\perp }}^{T}$, $\mathcal{H}_{k_{\parallel
}k_{\perp }}^{T}$and $\overline{\mathcal{H}}_{k_{\parallel }k_{\perp }}^{T}$
. Nevertheless it is common knowledge that in general relativity
gravitational waves carry only two degrees of freedom, represented by the $+$
and $\times $ polarisations. In this subsection we address this apparent
mismatch in the degree of freedom counting.

We start by writing the geometrical optics limit of the uncoupled first
order equations (\ref{4}) and (\ref{5}). {\color{black}By} employing that at least one of
the contitions $k_{\parallel }\gg 1$ or $k_{\perp }\gg 1$ holds together
with $\mathcal{O}\left( a_{i}^{2}B\right) =\mathcal{O}\left(
a_{i}^{2}P\right) \gg \mathcal{O}\left( C\right) =\mathcal{O}\left( D\right)
=\mathcal{O}\left( a_{i}E\right) =\mathcal{O}\left( a_{i}F\right) =\mathcal{O
}\left( a_{i}^{-1}S\right) =\mathcal{O}\left( 1\right) $ {\color{black} and} also the estimates
(\ref{kpera}), (\ref{Bgeom}) and (\ref{CDEgeom}) we obtain
\begin{equation}
\dot{\overline{\mathcal{E}}}_{k_{\parallel }k_{\perp }}^{T}\!\!\!=\frac{
2ik_{\parallel }}{a_{1}}\left( \frac{k_{\parallel }^{2}}{a_{1}^{2}}+\frac{
k_{\perp }^{2}}{a_{2}^{2}}\right) \left( \frac{2k_{\parallel }^{2}}{a_{1}^{2}
}+\frac{k_{\perp }^{2}}{a_{2}^{2}}\right) ^{-1}\mathcal{H}_{k_{\parallel
}k_{\perp }}^{T},
\end{equation}
\begin{equation}
\dot{\mathcal{H}}_{k_{\parallel }k_{\perp }}^{T}=-\frac{a_{1}}{%
2ik_{\parallel }}\left( \frac{2k_{\parallel }^{2}}{a_{1}^{2}}+\frac{k_{\perp
}^{2}}{a_{2}^{2}}\right) \overline{\mathcal{E}}_{k_{\parallel }k_{\perp
}}^{T}\ ,
\end{equation}
Hence we have found that $\overline{\mathcal{E}}_{k_{\parallel }k_{\perp
}}^{T}$ and $\mathcal{H}_{k_{\parallel }k_{\perp }}^{T}$ are simply related,
they represent the same degree of freedom. We have already shown that to
leading order they obey undampened wave equations with the propagation speed
of light. Again, in the geometrical optics limit the prefactors of the right
hand sides can be considered constants, hence $\overline{\mathcal{E}}
_{k_{\parallel }k_{\perp }}^{T}$ and $\mathcal{H}_{k_{\parallel }k_{\perp
}}^{T}$ are simply the time derivatives of each other, representing the same
gravitational degree of freedom. Note that this analysis could have been done 
also for the general case in section \ref{4.3.1}, where the same conclusions can be 
drawn from the system (\ref{4}-\ref{5}). However, since the couplings for the system 
in section \ref{4.3.2} are more intricate, we have chosen to work in the geometrical optics 
limit throughout in this section.

Next we revisit the geometrical optics limit of the coupled first order
equations (\ref{eq13}) and (\ref{eq14}), which simplify as
\begin{equation}
\dot{\mathcal{E}}_{k_{\parallel }k_{\perp }}^{T}\!=\frac{a_{1}}{
2ik_{\parallel }}\left( \frac{2k_{\parallel }^{2}}{a_{1}^{2}}+\frac{k_{\perp
}^{2}}{a_{2}^{2}}\right) \overline{\mathcal{H}}_{k_{\parallel }k_{\perp
}}^{T}\ ,
\end{equation}
\begin{equation}
\dot{\overline{\mathcal{H}}}_{k_{\parallel }k_{\perp }}^{T}\!\!\!\!=-\frac{
2ik_{\parallel }}{a_{1}}\left( \frac{k_{\parallel }^{2}}{a_{1}^{2}}+\frac{
k_{\perp }^{2}}{a_{2}^{2}}\right) \left( \frac{2k_{\parallel }^{2}}{a_{1}^{2}
}+\frac{k_{\perp }^{2}}{a_{2}^{2}}\right) ^{-1}\mathcal{E}_{k_{\parallel
}k_{\perp }}^{T}\ .
\end{equation}
Thus, again, the prefactors on the right hand sides can be considered
constants in the geometrical optics approximation, thus the Weyl variables $
\mathcal{E}_{k_{\parallel }k_{\perp }}^{T}$ and $\overline{\mathcal{H}}
_{k_{\parallel }k_{\perp }}^{T}$satisfying undampened wave equations with
the propagation speed of light are the time derivatives of each other,
representing the same gravitational degree of freedom. Note that in the
geometrical optics limit this second gravitational degree of freedom also
decoupled from matter and all four quantities $\mathcal{Y}=\left\{ \mathcal{E
}_{k_{\parallel }k_{\perp }}^{T},\overline{\mathcal{E}}_{k_{\parallel
}k_{\perp }}^{T},\mathcal{H}_{k_{\parallel }k_{\perp }}^{T},\overline{
\mathcal{H}}_{k_{\parallel }k_{\perp }}^{T}\right\} $ obey 
\begin{equation}
\mathcal{\ddot{Y}}+\left( \frac{k_{\parallel }^{2}}{a_{1}^{2}}+\frac{
k_{\perp }^{2}}{a_{2}^{2}}\right) \mathcal{Y}=0~,
\end{equation}
but they represent only two degrees of freedom.
}

{\color{black}{Like in the case of FLRW perturbations there are two matter degrees of freedom. One for the fluctuations of the density (via gradients of the energy conservation equation) and the other coming from the scalar part of the shear equation (representing velocity perturbations). }}

\section{Concluding Remarks\label{Concl}}


A general treatment of vorticity-free, perfect fluid perturbations of
Kantowski-Sachs models with a positive cosmological constant was considered
within the framework of the 1+1+2 covariant decomposition of spacetime. We
showed that the system of perturbation equations can be organised into a
hierarchy of three systems, namely (i) two coupled gravito-magnetic first
order differential equations, (ii) four first order differential equations
for the two complementary gravito-magnetic variables, a variable describing
the shear of the world-lines and the gradient of the matter density
perturbation, (iii) an extended set of algebraic relations involving all
variables, which provides a way of determining their evolution.

By assuming that the perturbation wavelengths is much smaller than the
curvature radius of the Kantowski-Sachs background, we were able to use the
geometrical optics approximation to describe the evolution of high frequency
perturbations. We found that system (i) gave rise to the leading order
decoupled propagation equations for gravitational waves on this background,
while to the next order, damping effects make the propagation along the
spheres dephased. At leading order, system (ii) gives rise to two decoupled
gravitational wave propagation equations for the complementary
gravito-magnetic variables, supplemented by wavelike evolutions for both the
shear and matter gradient perturbations, which both propagate with the same
speed of sound $c_{s}<1$, {\color{black} out of phase by} $\pi /2$. At the
next order the gravito-magnetic oscillations are again damped, while the
shear and matter waves obey forced oscillation wave equations. 

We note that the perfect fluid is marginally stable under the
vorticity-free, anisotropic pressure-avoiding perturbations at high
frequency. The degrees of freedom propagating as gravitational waves in the
geometrical optics approximation are exactly the even and odd tensorial
perturbations of {\color{black} both} the electric and magnetic parts of the Weyl tensor, in
agreement with its generic interpretation. {\color{black} While we have
found four such quantities obeying undampened wave equations with the
propagation speed of light, the even electric and odd magnetic Weyl
projections $\left( \mathcal{E}_{k_{\parallel }k_{\perp }}^{T},\overline{%
\mathcal{H}} _{k_{\parallel }k_{\perp }}^{T}\right) $ represent the same
gravitational degree of freedom, while the odd electric and even magnetic
Weyl projections $\left( \overline{\mathcal{E}}_{k_{\parallel }k_{\perp
}}^{T},\mathcal{H}_{k_{\parallel }k_{\perp }}^{T}\right) $ the other one. }

Beyond the geometrical optics approximation we have found indications for
the existence of direction dependent dispersion relations. {\color{black}
Remarkably, the second gravitational degree of freedom $\left( \overline{ 
\mathcal{E}}_{k_{\parallel }k_{\perp }}^{T},\mathcal{H}_{k_{\parallel
}k_{\perp }}^{T}\right) $ does not decouple from the matter density
perturbation, unlike in Friedman universes.}

\section*{ACKNOWLEDGMENTS}

Z.K. was supported by Hungarian Scientific Research Fund - OTKA Grant No. 100216. PKSD is supported by
the National Research Foundation (South Africa).

\appendix

\section{\textcolor{black}{The relation between the 2D and
4D
curvature
tensors\label{2Dcurvtens}}}

\textcolor{black}{We give here the proof for the equivalency of Eqs. (\ref{2Dcurvtensdef}) and
(\ref{2Dcurvtensdef2}). The second covariant derivative of any 2D dual
vector field $V_{k}$ projected to the 2D subspace with $N_{a}^{\,\,\,i}N_{b}^{\,\,\,j}N_{c}^{\,\,\,k}$ gives\begin{eqnarray}
N_{a}^{\,\,\,i}N_{b}^{\,\,\,j}N_{c}^{\,\,\,k}\nabla _{i}\nabla _{j}V_{k}
&=&\delta _{a}\delta _{b}V_{c}+V^{l}\left[ \left( \delta _{b}u_{l}\right)
\left( \delta _{a}u_{c}\right) -\left( \delta _{b}n_{l}\right) \left( \delta
_{a}n_{c}\right) \right]   \nonumber \\
&&-\left( \delta _{a}u_{b}\right) \dot{V}_{\bar{c}}+\left( \delta
_{a}n_{b}\right) \widehat{V}_{\bar{c}}~.
\end{eqnarray}Then from the definition of the Riemann tensor $\left( \nabla _{i}\nabla
_{j}-\nabla _{j}\nabla _{i}\right) V_{k}=R_{ijkl}V^{l}$, we find\begin{eqnarray}
N_{a}^{\,\,\,i}N_{b}^{\,\,\,j}N_{c}^{\,\,\,k}R_{ijkl}V^{l} &=&2\delta
_{\lbrack a}\delta _{b]}V_{c}-2\left( \delta _{\lbrack a}u_{b]}\right) \dot{V}_{\bar{c}}+2\left( \delta _{\lbrack a}n_{b]}\right) \widehat{V}_{\bar{c}} 
\nonumber \\
&&-\left[ \left( \delta _{a}u_{l}\right) \left( \delta _{b}u_{c}\right)
-\left( \delta _{a}u_{c}\right) \left( \delta _{b}u_{l}\right) \right.  
\nonumber \\
&&\left. -\left( \delta _{a}n_{l}\right) \left( \delta _{b}n_{c}\right)
+\left( \delta _{a}n_{c}\right) \left( \delta _{b}n_{l}\right) \right]
V^{l}~.
\end{eqnarray}With $\delta _{\lbrack a}u_{b]}=\Omega \varepsilon _{ab}$ and $\delta
_{\lbrack a}n_{b]}=\xi \varepsilon _{ab}$, the above identity reduces to\begin{eqnarray}
\delta _{\lbrack a}\delta _{b]}V_{c}-\Omega \varepsilon _{ab}\dot{V}_{\bar{c}}+\xi \varepsilon _{ab}\widehat{V}_{\bar{c}} &=&\frac{V^{d}}{2}\left[
N_{a}^{\,\,\,i}N_{b}^{\,\,\,j}N_{c}^{\,\,\,k}N_{d}^{\,\,\,l}R_{ijkl}+\left(
\delta _{a}u_{d}\right) \left( \delta _{b}u_{c}\right) \right.   \nonumber \\
&&\left. -\left( \delta _{a}u_{c}\right) \left( \delta _{b}u_{d}\right)
-\left( \delta _{a}n_{d}\right) \left( \delta _{b}n_{c}\right) +\left(
\delta _{a}n_{c}\right) \left( \delta _{b}n_{d}\right) \right] ~.
\end{eqnarray}The square bracket on the right hand side is the 2D curvature tensor $\mathcal{R}_{abcd}$, as can be seen by comparing the left hand side with the
definition (\ref{2Dcurvtensdef}).}

\section{Infinitesimal frame transformations on the Kantowski-Sachs background filled with perfect fluid\label{Frame}}

An infinitesimal frame transformation from the dyad ($u^{a}$, $n^{a}$) to
the dyad ($\overline{u}^{a}$, $\overline{n}^{a}$) can be defined as (see for
higher dimensional spacetime \cite{3+1+1}):%
\begin{eqnarray}
\overline{u}_{a} &=&u_{a}+\upsilon _{a}+\nu n_{a}\ ,\ \text{with\ \ }%
u^{a}\upsilon _{a}=n^{a}\upsilon _{a}=0,  \label{ua_frame_change} \\
\overline{n}_{a} &=&n_{a}+l_{a}+mu_{a}\ ,\ \text{with\ \ }%
u^{a}l_{a}=n^{a}l_{a}=0,  \label{na_frame_change}
\end{eqnarray}%
where $\upsilon _{a},~l_{a},~\nu ,~m$ are all first order. We will neglect
the second order contributions. The new dyad also obeys 
\begin{equation}
\overline{u}^{a}\overline{u}_{a}=-1\ ,\ \overline{n}^{a}\overline{n}_{a}=1\
,\ \overline{u}^{a}\overline{n}_{a}=0\ ,
\end{equation}%
which implies%
\begin{equation}
\nu =m\ .
\end{equation}%
There are five gauge degrees of freedom to fix the frame on the perturbed
spacetime. The transformations with $\upsilon _{a}=l_{a}=0$ represent 2D
infinitesimal Lorentz boosts, while the parameters $\upsilon _{a}$ and $%
l_{a} $ are related to infinitesimal translations.

The fundamental algebraic tensors $N_{ab}$ and $\varepsilon _{ab}$ change
accordingly:%
\begin{eqnarray}
\overline{N}_{ab} &=&N_{ab}+2u_{(a}\upsilon _{b)}-2n_{(a}l_{b)}\ ,
\label{Nbar} \\
\overline{\varepsilon }_{ab} &=&\varepsilon _{ab}+2n_{[a}\varepsilon
_{b]c}l^{c}-2u_{[a}\varepsilon _{b]c}\upsilon ^{c}\ .
\end{eqnarray}%
The new 2-metric obeys $\overline{N}_{ab}\overline{n}^{a}=\overline{N}_{ab}%
\overline{u}^{a}=0$.

The kinematic quantities defined for the new dyad vectors arise from the
decomposition of the covariant derivatives of $\overline{u}_{a}$ and $%
\overline{n}_{a}$ similarly to that given in Section \ref{decomp_sec}. This
implies the following transformations rules on Kantowski-Sachs background
for the kinematic quantities:%
\begin{equation}
\overline{\Theta }=\Theta +\widehat{\nu }+\delta _{a}\upsilon ^{a}\ ,
\end{equation}%
\begin{equation}
\overline{\mathcal{A}}=\mathcal{A}+\dot{\nu}+\left( \Sigma +\frac{\Theta }{3}%
\right) \nu \ ,
\end{equation}%
\begin{equation}
\overline{\Omega }=\Omega +\frac{1}{2}\varepsilon ^{ab}\delta _{a}\upsilon
_{b}\ ,
\end{equation}%
\begin{equation}
\overline{\Sigma }=\Sigma +\frac{2}{3}\widehat{\nu }-\frac{1}{3}\delta
_{a}\upsilon ^{a}\ ,
\end{equation}%
\begin{equation}
\overline{\phi }=\phi +\delta _{a}l^{a}-\left( \Sigma -\frac{2\Theta }{3}%
\right) \nu \ ,
\end{equation}%
\begin{equation}
\overline{\xi }=\xi +\varepsilon ^{ab}\delta _{a}l_{b}\ ,
\end{equation}%
\begin{equation}
\overline{\mathcal{A}}_{a}=\mathcal{A}_{a}+\dot{\upsilon}_{\bar{a}}-\frac{1}{%
2}\left( \Sigma -\frac{2\Theta }{3}\right) \upsilon _{a}\ ,
\end{equation}%
\begin{equation}
\overline{\Omega }_{a}=\Omega _{a}-\frac{1}{2}\varepsilon _{ab}\left( 
\widehat{\upsilon }^{b}-\delta ^{b}\nu \right) \ ,
\end{equation}%
\begin{equation}
\overline{\Sigma }_{a}=\Sigma _{a}+\frac{1}{2}\left( \widehat{\upsilon }_{%
\bar{a}}+\delta _{a}\nu -3\Sigma l_{a}\right) \ ,
\end{equation}%
\begin{equation}
\overline{a}_{a}=a_{a}+\widehat{l}_{\bar{a}}-\left( \Sigma +\frac{\Theta }{3}%
\right) \upsilon _{a}\ ,
\end{equation}%
\begin{equation}
\overline{\alpha }_{a}=\alpha _{a}+\dot{l}_{\bar{a}}\ ,
\end{equation}%
\begin{equation}
\overline{\Sigma }_{ab}=\Sigma _{ab}+\delta _{\{a}\upsilon _{b\}}\ ,
\end{equation}%
\begin{equation}
\overline{\zeta }_{ab}=\zeta _{ab}+\delta _{\{a}l_{b\}}\ ,
\end{equation}

The gravito-electro-magnetic quantities $\mathcal{E}$, $\mathcal{H}$, $%
\mathcal{E}_{ab}$ and $\mathcal{H}_{ab}$\ are invariant under the
infinitesimal frame change, while the transformation laws of $\mathcal{E}%
_{a} $ and $\mathcal{H}_{a}$ are 
\begin{equation}
\overline{\mathcal{E}}_{a}=\mathcal{E}_{a}-\frac{3\mathcal{E}}{2}l_{c}\ ,
\end{equation}%
\begin{equation}
\overline{\mathcal{H}}_{a}=\mathcal{H}_{a}-\frac{3\mathcal{E}}{2}\varepsilon
_{ab}\upsilon ^{b}\ .
\end{equation}

The matter variables $\mu $, $p$, $\Pi $, $\Pi _{a}$ and $\Pi _{ab}$ are
invariant under the infinitesimal frame change, while $Q$ and $Q_{a}$
describing the energy current transform as%
\begin{equation}
\overline{Q}=Q-\left( \mu +p\right) \nu \ ,
\end{equation}%
\begin{equation}
\overline{Q}_{a}=Q_{a}-\left( \mu +p\right) \upsilon _{a}\ .
\end{equation}

The gauge-invariant variables defined by Eq. (\ref{gaugeinv}) transform as%
\begin{equation}
G_{a}=G_{a}+\dot{G}\upsilon _{a}\ ,
\end{equation}%
where $G_{a}\equiv \left\{ \mu _{a},X_{a},V_{a},W_{a},p_{a}\right\} $ and $%
G\equiv \left\{ \mu ,\mathcal{E},\Sigma ,\Theta ,p\right\} $, respectively.

\section{Commutation relations\label{commutation}}

The commutation relations of covariant derivatives of the scalar field $\Psi 
$ on Kantowski-Sachs background, to first order are%
\begin{equation}
\widehat{\dot{\Psi}}-\dot{\widehat{\Psi }}=-\mathcal{A}\dot{\Psi}+\left(
\Sigma +\frac{\Theta }{3}\right) \widehat{\Psi }\ ,
\end{equation}%
\begin{equation}
\delta _{a}\dot{\Psi}-N_{a}^{\,\,\,b}\left( \delta _{b}\Psi \right) ^{\cdot
}=-\mathcal{A}_{a}\dot{\Psi}-\!\!\frac{1}{2}\left( \Sigma -\frac{2\Theta }{3}%
\right) \delta _{a}\Psi \ ,
\end{equation}%
\begin{equation}
\delta _{a}\widehat{\Psi }-N_{a}^{\,\,\,b}\left( \widehat{\delta _{b}\Psi }%
\right) =-2\varepsilon _{ab}\Omega ^{b}\dot{\Psi}\ ,  \label{scalcomm3}
\end{equation}%
\begin{equation}
\delta _{\lbrack a}\delta _{b]}\Psi =\varepsilon _{ab}\Omega \dot{\Psi}\ .
\label{scalcomm4}
\end{equation}

Similar relations hold for the first order 2-vector $\Psi _{a}$:%
\begin{equation}
\widehat{\dot{\Psi}}_{\bar{a}}-\dot{\widehat{\Psi }}_{\bar{a}}=\left( \Sigma
+\frac{\Theta }{3}\right) \widehat{\Psi }_{\bar{a}}\ ,
\end{equation}%
\begin{equation}
\delta _{a}\dot{\Psi}_{b}-N_{a}^{\,\,\,c}N_{b}^{\,\,\,d}\left( \delta
_{c}\Psi _{d}\right) ^{\cdot }=-\frac{1}{2}\left( \Sigma -\frac{2\Theta }{3}%
\right) \delta _{a}\Psi _{b}\ ,
\end{equation}%
\begin{equation}
\delta _{a}\widehat{\Psi }_{b}-N_{a}^{\,\,\,c}N_{b}^{\,\,\,d}\left( \widehat{%
\delta _{c}\Psi _{d}}\right) =0\ ,
\end{equation}%
\begin{equation}
\delta _{\lbrack a}\delta _{b]}\Psi _{c}=2\mathcal{R}N_{c[a}\Psi _{b]}\ ,
\end{equation}%
and for the first order symmetric, trace-free 2-tensor $\Psi _{ab}$:%
\begin{equation}
\widehat{\dot{\Psi}}_{\{ab\}}-\dot{\widehat{\Psi }}_{\{ab\}}=\left( \Sigma +%
\frac{\Theta }{3}\right) \widehat{\Psi }_{\bar{a}}\ ,
\end{equation}%
\begin{equation}
\delta _{a}\dot{\Psi}_{bc}-N_{a}^{\,\,\,d}N_{b}^{\,\,\,e}N_{c}^{\,\,\,f}%
\left( \delta _{d}\Psi _{ef}\right) ^{\cdot }=-\frac{1}{2}\left( \Sigma -%
\frac{2\Theta }{3}\right) \delta _{a}\Psi _{bc}\ ,
\end{equation}%
\begin{equation}
\delta _{a}\widehat{\Psi }_{bc}-N_{a}^{\,\,\,d}N_{b}^{\,\,\,e}N_{c}^{\,\,%
\,f}\left( \widehat{\delta _{d}\Psi _{ef}}\right) =0\ ,
\end{equation}%
\begin{equation}
2\delta _{\lbrack a}\delta _{b]}\Psi _{cd}=\mathcal{R}\left( N_{c[a}\Psi
_{b]d}+N_{d[a}\Psi _{b]c}\right) \ .
\end{equation}%
where $\mathcal{R}$ is given by Eq. (\ref{2TimesGaussianCurv}).

\section{Properties of vector and tensor spherical harmonics\label{harmonics}%
}

In this Appendix we enlist a set of identities for the even $Q_{a}^{k_{\perp
}}$ and odd $\overline{Q}_{a}^{k_{\perp }}$ vector spherical harmonics,
including the orthogonality relations%
\begin{equation}
N^{ab}Q_{a}^{k_{\perp }}\overline{Q}_{b}^{k_{\perp }}=0~,
\end{equation}%
the algebraic relations%
\begin{equation}
Q_{a}^{k_{\perp }}=-\varepsilon _{a}^{\,\,\,\,b}\overline{Q}_{b}^{k_{\perp
}}\ ,\quad \overline{Q}_{a}^{k_{\perp }}=\varepsilon
_{a}^{\,\,\,\,b}Q_{b}^{k_{\perp }}\ ,\ \ 
\end{equation}%
and the differential relations%
\begin{equation}
\dot{Q}_{a}^{k_{\perp }}=\widehat{Q}_{a}^{k_{\perp }}=0\ ,\quad \dot{%
\overline{Q}}_{a}^{k_{\perp }}=\widehat{\overline{Q}}_{a}^{k_{\perp }}=0\ ,
\end{equation}%
\begin{equation}
\delta ^{2}Q_{a}^{k_{\perp }}=\frac{1-k_{\perp }^{2}}{a_{2}^{2}}%
Q_{a}^{k_{\perp }}\ ,\quad \delta ^{2}\overline{Q}_{a}^{k_{\perp }}=\frac{%
1-k_{\perp }^{2}}{a_{2}^{2}}\overline{Q}_{a}^{k_{\perp }}\ ,
\end{equation}%
\begin{equation}
\delta ^{a}Q_{a}^{k_{\perp }}=-\frac{k_{\perp }^{2}}{a_{2}}Q^{k_{\perp }}\
,\quad \delta ^{a}\overline{Q}_{a}^{k_{\perp }}=0\ ,  \label{vectspherid3}
\end{equation}%
\begin{equation}
\varepsilon ^{ab}\delta _{a}Q_{b}^{k_{\perp }}=0\ ,\quad \varepsilon
^{ab}\delta _{a}\overline{Q}_{b}^{k_{\perp }}=\frac{k_{\perp }^{2}}{a_{2}}%
Q^{k_{\perp }}~.  \label{vectspherid4}
\end{equation}

The even and odd tensor spherical harmonics obey in turn the orthogonality
relations%
\begin{equation}
N^{ab}N^{cd}Q_{ac}^{k_{\perp }}\overline{Q}_{bd}^{k_{\perp }}=0~,
\end{equation}%
the algebraic relations%
\begin{equation}
Q_{ab}^{k_{\perp }}=\varepsilon _{\{a}^{\,\,\,\,\,\,\,c}\overline{Q}%
_{b\}c}^{k_{\perp }}\ ,\quad \overline{Q}_{ab}^{k_{\perp }}=-\varepsilon
_{\{a}^{\,\,\,\,\,\,\,c}Q_{b\}c}^{k_{\perp }}\ ,\ \ 
\end{equation}%
and the differential relations%
\begin{equation}
\dot{Q}_{ab}^{k_{\perp }}=\widehat{Q}_{ab}^{k_{\perp }}=0\ ,\quad \dot{%
\overline{Q}}_{ab}^{k_{\perp }}=\widehat{\overline{Q}}_{ab}^{k_{\perp }}=0\ ,
\end{equation}%
\begin{equation}
\delta ^{2}Q_{ab}^{k_{\perp }}=\frac{4-k_{\perp }^{2}}{a_{2}^{2}}%
Q_{ab}^{k_{\perp }}\ ,\quad \delta ^{2}\overline{Q}_{ab}^{k_{\perp }}=\frac{%
4-k_{\perp }^{2}}{a_{2}^{2}}\overline{Q}_{ab}^{k_{\perp }}~,
\end{equation}%
\begin{equation}
\delta ^{b}Q_{ab}^{k_{\perp }}=\frac{2-k_{\perp }^{2}}{2a_{2}}%
Q_{a}^{k_{\perp }}\ ,\quad \delta ^{b}\overline{Q}_{ab}^{k_{\perp }}=-\frac{%
2-k_{\perp }^{2}}{2a_{2}}\overline{Q}_{a}^{k_{\perp }}~,
\end{equation}%
\begin{equation}
\varepsilon _{a}^{\,\,\,\,\,c}\delta ^{b}Q_{bc}^{k_{\perp }}=\frac{%
2-k_{\perp }^{2}}{2a_{2}}\overline{Q}_{a}^{k_{\perp }}\ ,\quad \varepsilon
_{a}^{\,\,\,\,\,c}\delta ^{b}\overline{Q}_{bc}^{k_{\perp }}=\frac{2-k_{\perp
}^{2}}{2a_{2}}Q_{a}^{k_{\perp }}~,
\end{equation}%
\begin{equation}
\varepsilon ^{bc}\delta _{b}Q_{ac}^{k_{\perp }}=\frac{2-k_{\perp }^{2}}{%
2a_{2}}\overline{Q}_{a}^{k_{\perp }}\ ,\quad \varepsilon ^{bc}\delta _{b}%
\overline{Q}_{ac}^{k_{\perp }}=\frac{2-k_{\perp }^{2}}{2a_{2}}%
Q_{a}^{k_{\perp }}~.
\end{equation}

\noindent 
\textcolor{black}{For reviews of various types of harmonics used in
relativity see, e.g., \cite{Thorne, Harrison}.}

\section{Harmonic expansion of the vorticity-free perturbation equations 
\label{perturbEq_harmonics}}

We give here the harmonic decomposition of Eqs. (\ref{phidot})-(\ref{constr3}%
), employing Eq. (\ref{OmegaSrel}) and $\overline{\mathcal{A}}_{k_{\parallel
}k_{\perp }}^{V}=0$. The perturbation equations decouple into two sets,
describing the even and odd parity sectors.

\subsection{Odd parity sector}

The evolution equations for gauge-invariant 2-vector perturbation variables
with odd parity are%
\begin{equation}
\dot{\overline{a}}_{k_{\parallel }k_{\perp }}^{V}\!=\frac{ik_{\parallel }}{%
a_{1}}\overline{\alpha }_{k_{\parallel }k_{\perp }}^{V}\!\!\!\!+\mathcal{H}%
_{k_{\parallel }k_{\perp }}^{V}-\!\!\left( \Sigma \!+\!\frac{\Theta }{3}%
\right) \!\overline{a}_{k_{\parallel }k_{\perp }}^{V}\!\!\ ,  \label{var7}
\end{equation}%
\begin{equation}
2\dot{\overline{\Sigma }}_{k_{\parallel }k_{\perp }}^{V}=-\left( \Sigma +%
\frac{4\Theta }{3}\right) \overline{\Sigma }_{k_{\parallel }k_{\perp
}}^{V}-3\Sigma \overline{\alpha }_{k_{\parallel }k_{\perp }}^{V}-2\overline{%
\mathcal{E}}_{k_{\parallel }k_{\perp }}^{V},
\end{equation}%
\begin{eqnarray}
\dot{\overline{\mathcal{E}}}_{k_{\parallel }k_{\perp }}^{V}\!\! &=&-\frac{%
ik_{\parallel }}{2a_{1}}\mathcal{H}_{k_{\parallel }k_{\perp }}^{V}+\!\frac{3%
}{4}\!\left( \!\Sigma \!-\!\frac{4\Theta }{3}\!\right) \overline{\mathcal{E}}%
_{k_{\parallel }k_{\perp }}^{V}\!\!\!\!+\frac{3}{4a_{2}}\mathcal{H}%
_{k_{\parallel }k_{\perp }}^{S}  \nonumber \\
&&+\frac{\left( 3\mathcal{E\!}-\!2\mu \!-\!2p\right) }{4}\overline{\Sigma }%
_{k_{\parallel }k_{\perp }}^{V}\!\!+\frac{2-k_{\perp }^{2}}{4a_{2}}\mathcal{H%
}_{k_{\parallel }k_{\perp }}^{T}-\!\frac{3\mathcal{E}}{2}\overline{\alpha }%
_{k_{\parallel }k_{\perp }}^{V},
\end{eqnarray}%
\begin{equation}
\dot{\overline{\mathcal{H}}}_{k_{\parallel }k_{\perp }}^{V}\!\!\!\!\!=\frac{%
ik_{\parallel }}{2a_{1}}\mathcal{E}_{k_{\parallel }k_{\perp }}^{V}+\frac{3}{4%
}\!\!\left( \Sigma -\frac{4\Theta }{3}\right) \overline{\mathcal{H}}%
_{k_{\parallel }k_{\perp }}^{V}-\frac{3}{4}X_{k_{\parallel }k_{\perp }}^{V}-%
\frac{3\mathcal{E}}{2}\mathcal{A}_{k_{\parallel }k_{\perp }}^{V}\!\!+\frac{3%
\mathcal{E}}{4}a_{k_{\parallel }k_{\perp }}^{V}\!\!-\frac{2-k_{\perp }^{2}}{%
4a_{2}}\mathcal{E}_{k_{\parallel }k_{\perp }}^{T},
\end{equation}%
The evolution equations for the odd parity tensor perturbations are%
\begin{equation}
\dot{\overline{\Sigma }}_{k_{\parallel }k_{\perp }}^{T}\!=\!\!\left(
\!\Sigma \!-\!\frac{2\Theta }{3}\!\right) \!\overline{\Sigma }_{k_{\parallel
}k_{\perp }}^{T}\!\!\!\!-\!\overline{\mathcal{E}}_{k_{\parallel }k_{\perp
}}^{T},
\end{equation}%
\begin{equation}
\dot{\overline{\zeta }}_{k_{\parallel }k_{\perp }}^{T}=\frac{1}{2}\left(
\Sigma -\frac{2\Theta }{3}\right) \overline{\zeta }_{k_{\parallel }k_{\perp
}}^{T}-\frac{\overline{\alpha }_{k_{\parallel }k_{\perp }}^{V}}{a_{2}}-%
\mathcal{H}_{k_{\parallel }k_{\perp }}^{T}\ ,
\end{equation}%
\begin{equation}
\dot{\overline{\mathcal{E}}}_{k_{\parallel }k_{\perp }}^{T}\!\!\!\!\!\!=%
\frac{ik_{\parallel }}{a_{1}}\mathcal{H}_{k_{\parallel }k_{\perp }}^{T}-%
\frac{3}{2}\!\left( \!\Sigma \!+\!\frac{2\Theta }{3}\!\right) \!\overline{%
\mathcal{E}}_{k_{\parallel }k_{\perp }}^{T}\!\!\!-\!\frac{1}{a_{2}}\mathcal{H%
}_{k_{\parallel }k_{\perp }}^{V}-\frac{\left( 3\mathcal{E}+\mu +p\right) }{2}%
\overline{\Sigma }_{k_{\parallel }k_{\perp }}^{T},  \label{Eq4}
\end{equation}%
\begin{equation}
\dot{\overline{\mathcal{H}}}_{k_{\parallel }k_{\perp }}^{T}\!\!\!\!=-\frac{%
ik_{\parallel }}{a_{1}}\mathcal{E}_{k_{\parallel }k_{\perp }}^{T}-\frac{3}{2}%
\left( \Sigma +\frac{2\Theta }{3}\right) \overline{\mathcal{H}}%
_{k_{\parallel }k_{\perp }}^{T}+\frac{3\mathcal{E}}{2}\zeta _{k_{\parallel
}k_{\perp }}^{T}+\frac{1}{a_{2}}\mathcal{E}_{k_{\parallel }k_{\perp }}^{V}.
\label{Eq6}
\end{equation}%
They obey the constraints$\!\!\!$%
\begin{equation}
\!\!\!\frac{\mathcal{A}_{k_{\parallel }k_{\perp }}^{S}}{a_{2}}=\!\frac{%
ik_{\parallel }}{a_{1}}\mathcal{A}_{k_{\parallel }k_{\perp }}^{V}\!\!\ ,
\label{AvAs}
\end{equation}%
\begin{equation}
\frac{ik_{\parallel }}{a_{1}}\overline{\Sigma }_{k_{\parallel }k_{\perp
}}^{V}\!\!\!\!\!\!=-\frac{3\Sigma }{2}\overline{a}_{k_{\parallel }k_{\perp
}}^{V}\!\!+\!\frac{2-k_{\perp }^{2}}{2a_{2}}\overline{\Sigma }_{k_{\parallel
}k_{\perp }}^{T},  \label{SigmabarV1}
\end{equation}%
\begin{equation}
\!\frac{2ik_{\parallel }}{a_{1}}\overline{\mathcal{E}}_{k_{\parallel
}k_{\perp }}^{V}=-3\mathcal{E}\overline{a}_{k_{\parallel }k_{\perp
}}^{V}\!\!-3\Sigma \mathcal{H}_{k_{\parallel }k_{\perp }}^{V}+\frac{%
2-k_{\perp }^{2}}{a_{2}}\overline{\mathcal{E}}_{k_{\parallel }k_{\perp
}}^{T}\ ,  \label{epsbarV1}
\end{equation}%
\begin{equation}
\frac{2ik_{\parallel }}{a_{1}}\overline{\mathcal{H}}_{k_{\parallel }k_{\perp
}}^{V}\!\!\!=-3\mathcal{E}\Sigma _{k_{\parallel }k_{\perp }}^{V}+\frac{%
2-k_{\perp }^{2}}{a_{2}}\overline{\mathcal{H}}_{k_{\parallel }k_{\perp
}}^{T}+3\Sigma \mathcal{E}_{k_{\parallel }k_{\perp }}^{V}\!\!\ ,
\label{HbarV1}
\end{equation}%
\begin{equation}
\frac{ik_{\parallel }}{a_{1}}\overline{\Sigma }_{k_{\parallel }k_{\perp
}}^{T}\!\!=-\frac{\overline{\Sigma }_{k_{\parallel }k_{\perp }}^{V}}{a_{2}}+%
\frac{3\Sigma }{2}\overline{\zeta }_{k_{\parallel }k_{\perp }}^{T}\!\!-%
\mathcal{H}_{k_{\parallel }k_{\perp }}^{T}\!,  \label{SigmabarT}
\end{equation}%
\begin{equation}
\frac{ik_{\parallel }}{a_{1}}\overline{\zeta }_{k_{\parallel }k_{\perp
}}^{T}=\left( \Sigma +\frac{\Theta }{3}\right) \overline{\Sigma }%
_{k_{\parallel }k_{\perp }}^{T}-\frac{1}{a_{2}}\overline{a}_{k_{\parallel
}k_{\perp }}^{V}-\overline{\mathcal{E}}_{k_{\parallel }k_{\perp }}^{T},
\label{shearbarT1}
\end{equation}%
\begin{equation}
\frac{\xi _{k_{\parallel }k_{\perp }}^{S}}{a_{2}}-\frac{2\!-\!k_{\perp }^{2}%
}{2a_{2}}\overline{\zeta }_{k_{\parallel }k_{\perp }}^{T}=\!\!\left( \Sigma -%
\frac{2\Theta }{3}\right) \!\frac{\overline{\Sigma }_{k_{\parallel }k_{\perp
}}^{V}}{2}+\overline{\mathcal{E}}_{k_{\parallel }k_{\perp }}^{V},
\label{ksziS2}
\end{equation}%
\begin{equation}
2\mathcal{H}_{k_{\parallel }k_{\perp }}^{V}\!=\frac{2-k_{\perp }^{2}}{a_{2}}%
\overline{\Sigma }_{k_{\parallel }k_{\perp }}^{T}.  \label{HV}
\end{equation}

\subsection{Even parity sector}

The evolution equations governing the gauge-invariant scalar perturbation
variables with even parity are%
\begin{equation}
\dot{\phi}_{k_{\parallel }k_{\perp }}^{S}\!\!=\!\left( \!\Sigma \!-\!\frac{%
2\Theta }{3}\!\right) \!\!\left( \frac{\phi _{k_{\parallel }k_{\perp }}^{S}}{%
2}\!\!-\mathcal{A}_{k_{\parallel }k_{\perp }}^{S}\!\right) \!-\frac{k_{\perp
}^{2}}{a_{2}}\alpha _{k_{\parallel }k_{\perp }}^{V},  \label{phiSdot}
\end{equation}%
\begin{equation}
2\dot{\xi}_{k_{\parallel }k_{\perp }}^{S}=\left( \Sigma -\frac{2\Theta }{3}%
\right) \xi _{k_{\parallel }k_{\perp }}^{S}+\mathcal{H}_{k_{\parallel
}k_{\perp }}^{S}+\frac{k_{\perp }^{2}}{a_{2}}\overline{\alpha }%
_{k_{\parallel }k_{\perp }}^{V}\ ,
\end{equation}%
\begin{equation}
\dot{\mathcal{H}}_{k_{\parallel }k_{\perp }}^{S}=\frac{3}{2}\!\!\left(
\Sigma -\frac{2\Theta }{3}\right) \mathcal{H}_{k_{\parallel }k_{\perp }}^{S}-%
\frac{k_{\perp }^{2}}{a_{2}}\overline{\mathcal{E}}_{k_{\parallel }k_{\perp
}}^{V}-3\mathcal{E}\xi _{k_{\parallel }k_{\perp }}^{S}\ .
\end{equation}%
The evolution of even parity 2-vector perturbations are given by%
\begin{equation}
\dot{\mu}_{k_{\parallel }k_{\perp }}^{V}=\!\!\left( \frac{\Sigma }{2}-\frac{%
4\Theta }{3}\right) \mu _{k_{\parallel }k_{\perp }}^{V}-\left( \mu +p\right)
W_{k_{\parallel }k_{\perp }}^{V}-\Theta p_{k_{\parallel }k_{\perp }}^{V}+%
\dot{\mu}\mathcal{A}_{k_{\parallel }k_{\perp }}^{V},  \label{Eq1}
\end{equation}%
\begin{eqnarray}
\dot{X}_{k_{\parallel }k_{\perp }}^{V} &=&2\left( \Sigma -\frac{2\Theta }{3}%
\right) X_{k_{\parallel }k_{\perp }}^{V}-\frac{\mu +p}{2}V_{k_{\parallel
}k_{\perp }}^{V}+\frac{3\mathcal{E}}{2}\!\left( \!V_{k_{\parallel }k_{\perp
}}^{V}\!\!-\!\frac{2}{3}W_{k_{\parallel }k_{\perp }}^{V}\!\right)  \nonumber
\\
&&+\dot{\mathcal{E}}\mathcal{A}_{k_{\parallel }k_{\perp }}^{V}-\frac{\Sigma 
}{2}\left( \mu _{k_{\parallel }k_{\perp }}^{V}+p_{k_{\parallel }k_{\perp
}}^{V}\right) +\frac{k_{\perp }^{2}}{a_{2}^{2}}\overline{\mathcal{H}}%
_{k_{\parallel }k_{\perp }}^{V},
\end{eqnarray}%
\begin{eqnarray}
\!\!\dot{V}_{k_{\parallel }k_{\perp }}^{V}\!\!-\!\frac{2}{3}\dot{W}%
_{k_{\parallel }k_{\perp }}^{V}\!\!\! &=&\!\!\left( \!\dot{\Sigma}\!-\!\frac{%
2\dot{\Theta}}{3}\!\right) \mathcal{A}_{k_{\parallel }k_{\perp }}^{V}\!\!+\!%
\frac{k_{\perp }^{2}}{a_{2}^{2}}\mathcal{A}_{k_{\parallel }k_{\perp }}^{V}+%
\frac{3}{2}\left( \!\Sigma \!-\!\frac{2\Theta }{3}\!\right) \!\!  \nonumber
\\
&&\times \left( \!V_{k_{\parallel }k_{\perp }}^{V}\!\!-\!\frac{2}{3}%
W_{k_{\parallel }k_{\perp }}^{V}\!\right) +\frac{1}{3}\left( \!\mu
_{k_{\parallel }k_{\perp }}^{V}\!\!+3p_{k_{\parallel }k_{\perp
}}^{V}\!\right) \!\!-\!X_{k_{\parallel }k_{\perp }}^{V}\!,
\end{eqnarray}%
\begin{eqnarray}
\dot{W}_{k_{\parallel }k_{\perp }}^{V}\!\! &=&\frac{ik_{\parallel }}{%
a_{1}a_{2}}\mathcal{A}_{k_{\parallel }k_{\perp }}^{S}+\left( \!\frac{\Sigma
\!}{2}-\Theta \!\right) \!W_{k_{\parallel }k_{\perp }}^{V}\!\!+\dot{\Theta}%
\mathcal{A}_{k_{\parallel }k_{\perp }}^{V}  \nonumber \\
&&-\frac{1}{2}\left( \mu _{k_{\parallel }k_{\perp }}^{V}+3p_{k_{\parallel
}k_{\perp }}^{V}\right) -3\Sigma V_{k_{\parallel }k_{\perp }}^{V}-\frac{%
k_{\perp }^{2}}{a_{2}^{2}}\mathcal{A}_{k_{\parallel }k_{\perp }}^{V},
\end{eqnarray}%
\begin{equation}
\dot{a}_{k_{\parallel }k_{\perp }}^{V}=\frac{ik_{\parallel }}{a_{1}}\alpha
_{k_{\parallel }k_{\perp }}^{V}-\overline{\mathcal{H}}_{k_{\parallel
}k_{\perp }}^{V}-\left( \Sigma +\frac{\Theta }{3}\right) \left( \mathcal{A}%
_{k_{\parallel }k_{\perp }}^{V}+a_{k_{\parallel }k_{\perp }}^{V}\right) ,
\label{var8}
\end{equation}%
\begin{equation}
2\dot{\Sigma}_{k_{\parallel }k_{\perp }}^{V}\!\!=\!\frac{ik_{\parallel }}{%
a_{1}}\mathcal{A}_{k_{\parallel }k_{\perp }}^{V}\!\!+\frac{1}{a_{2}}\mathcal{%
A}_{k_{\parallel }k_{\perp }}^{S}\!\!\!\!-3\Sigma \alpha _{k_{\parallel
}k_{\perp }}^{V}-\left( \Sigma +\frac{4\Theta }{3}\right) \Sigma
_{k_{\parallel }k_{\perp }}^{V}-2\mathcal{E}_{k_{\parallel }k_{\perp }}^{V}\
,  \label{SigmaVdot}
\end{equation}%
\begin{eqnarray}
\dot{\mathcal{E}}_{k_{\parallel }k_{\perp }}^{V} &=&\frac{3}{4}\left( \Sigma
-\frac{4\Theta }{3}\right) \mathcal{E}_{k_{\parallel }k_{\perp }}^{V}+\frac{%
\left( 3\mathcal{E\!}-\!2\mu \!-\!2p\right) }{4}\Sigma _{k_{\parallel
}k_{\perp }}^{V}  \nonumber \\
&&\!+\frac{ik_{\parallel }}{2a_{1}}\overline{\mathcal{H}}_{k_{\parallel
}k_{\perp }}^{V}-\frac{3\mathcal{E}}{2}\alpha _{k_{\parallel }k_{\perp
}}^{V}\!\!+\frac{2-k_{\perp }^{2}}{4a_{2}}\overline{\mathcal{H}}%
_{k_{\parallel }k_{\perp }}^{T}\ ,  \label{epsVdot}
\end{eqnarray}%
\begin{equation}
\dot{\mathcal{H}}_{k_{\parallel }k_{\perp }}^{V}\!\!\!\!=-\frac{%
ik_{\parallel }}{2a_{1}}\overline{\mathcal{E}}_{k_{\parallel }k_{\perp
}}^{V}+\frac{3}{4}\!\!\left( \Sigma -\frac{4\Theta }{3}\right) \mathcal{H}%
_{k_{\parallel }k_{\perp }}^{V}\!\!-\frac{3\mathcal{E}}{4}\overline{a}%
_{k_{\parallel }k_{\perp }}^{V}\!\!\!\!-\frac{2\!-\!k_{\perp }^{2}}{4a_{2}}%
\overline{\mathcal{E}}_{k_{\parallel }k_{\perp }}^{T}\!,
\end{equation}%
The evolution equations for the even parity 2-tensor perturbations are%
\begin{equation}
\dot{\Sigma}_{k_{\parallel }k_{\perp }}^{T}\!\!=\!\left( \!\Sigma \!-\!\frac{%
2\Theta }{3}\!\right) \Sigma _{k_{\parallel }k_{\perp }}^{T}\!\!+\frac{1}{%
a_{2}}\mathcal{A}_{k_{\parallel }k_{\perp }}^{V}\!\!-\!\mathcal{E}%
_{k_{\parallel }k_{\perp }}^{T},  \label{Eq2}
\end{equation}%
\begin{equation}
\dot{\zeta}_{k_{\parallel }k_{\perp }}^{T}\!\!=\!\frac{1}{2}\!\left(
\!\Sigma \!-\!\frac{2\Theta }{3}\!\right) \zeta _{k_{\parallel }k_{\perp
}}^{T}\!\!+\!\frac{1}{a_{2}}\alpha _{k_{\parallel }k_{\perp }}^{V}\!\!+\!%
\overline{\mathcal{H}}_{k_{\parallel }k_{\perp }}^{T},  \label{EqzetaTdot}
\end{equation}%
\begin{equation}
\dot{\mathcal{E}}_{k_{\parallel }k_{\perp }}^{T}\!=\!-\frac{ik_{\parallel }}{%
a_{1}}\overline{\mathcal{H}}_{k_{\parallel }k_{\perp }}^{T}\!-\frac{3}{2}%
\left( \Sigma +\frac{2\Theta }{3}\right) \mathcal{E}_{k_{\parallel }k_{\perp
}}^{T}-\frac{\left( 3\mathcal{E}+\mu +p\right) }{2}\Sigma _{k_{\parallel
}k_{\perp }}^{T}-\frac{1}{a_{2}}\overline{\mathcal{H}}_{k_{\parallel
}k_{\perp }}^{V},  \label{Eq3}
\end{equation}%
\begin{equation}
\dot{\mathcal{H}}_{k_{\parallel }k_{\perp }}^{T}=\frac{ik_{\parallel }}{a_{1}%
}\overline{\mathcal{E}}_{k_{\parallel }k_{\perp }}^{T}-\frac{3}{2}\left(
\Sigma +\frac{2\Theta }{3}\right) \mathcal{H}_{k_{\parallel }k_{\perp }}^{T}-%
\frac{3\mathcal{E}}{2}\overline{\zeta }_{k_{\parallel }k_{\perp }}^{T}+\frac{%
1}{a_{2}}\overline{\mathcal{E}}_{k_{\parallel }k_{\perp }}^{V}\ .
\label{Eq5}
\end{equation}

The constraint equations for the even parity sector are%
\begin{equation}
\!\!\!\!\frac{ik_{\parallel }}{a_{1}a_{2}}\phi _{k_{\parallel }k_{\perp
}}^{S}\!\!\!=\!\frac{1}{3}\!\left( \!\Sigma \!+\!\frac{4\Theta }{3}\!\right)
\!\!W_{k_{\parallel }k_{\perp }}^{V}\!\!-\!\!\left( \!2\Sigma \!-\!\frac{%
\Theta }{3}\!\right) \!V_{k_{\parallel }k_{\perp }}^{V}-\frac{k_{\perp }^{2}%
}{a_{2}^{2}}a_{k_{\parallel }k_{\perp }}^{V}-X_{k_{\parallel }k_{\perp
}}^{V}\!\!-\frac{2}{3}\mu _{k_{\parallel }k_{\perp }}^{V}\ ,  \label{phiS2}
\end{equation}%
\begin{equation}
\frac{ik_{\parallel }}{a_{1}}\xi _{k_{\parallel }k_{\perp }}^{S}=\frac{%
k_{\perp }^{2}}{2a_{2}}\overline{a}_{k_{\parallel }k_{\perp }}^{V}\ ,
\label{ksziS}
\end{equation}%
\begin{equation}
\frac{ik_{\parallel }}{a_{1}}\mathcal{H}_{k_{\parallel }k_{\perp }}^{S}\!\!=%
\frac{k_{\perp }^{2}}{a_{2}}\mathcal{H}_{k_{\parallel }k_{\perp }}^{V},
\label{HS}
\end{equation}%
\begin{equation}
\frac{ik_{\parallel }}{a_{1}}p_{k_{\parallel }k_{\perp }}^{V}\!\!=-\frac{\mu
+p}{a_{2}}\mathcal{A}_{k_{\parallel }k_{\perp }}^{S}\!\!\!\!\ ,  \label{pvAs}
\end{equation}%
\begin{equation}
\frac{ik_{\parallel }}{a_{1}}\!\!\left( X_{k_{\parallel }k_{\perp
}}^{V}\!\!-\!\frac{\mu _{k_{\parallel }k_{\perp }}^{V}}{3}\right) \!=\frac{%
k_{\perp }^{2}}{a_{2}^{2}}\mathcal{E}_{k_{\parallel }k_{\perp }}^{V}\!-\frac{%
3\mathcal{E}}{2a_{2}}\phi _{k_{\parallel }k_{\perp }}^{S}\!\!,  \label{Xmu}
\end{equation}%
\begin{equation}
\frac{ik_{\parallel }}{a_{1}}\!\!\left( V_{k_{\parallel }k_{\perp
}}^{V}\!\!-\!\frac{2}{3}W_{k_{\parallel }k_{\perp }}^{V}\!\right) =\frac{%
k_{\perp }^{2}}{a_{2}^{2}}\Sigma _{k_{\parallel }k_{\perp }}^{V}-\frac{%
3\Sigma }{2a_{2}}\phi _{k_{\parallel }k_{\perp }}^{S}\!\!\!\,,  \label{VW}
\end{equation}%
\begin{equation}
\!\!\!\frac{ik_{\parallel }}{a_{1}}\Sigma _{k_{\parallel }k_{\perp
}}^{V}\!\!=\!-\frac{3\Sigma }{2}a_{k_{\parallel }k_{\perp }}^{V}\!\!-\frac{%
2-k_{\perp }^{2}}{2a_{2}}\Sigma _{k_{\parallel }k_{\perp }}^{T}\!+\frac{1}{2}%
\!\left( V_{k_{\parallel }k_{\perp }}^{V}\!\!+\frac{4}{3}W_{k_{\parallel
}k_{\perp }}^{V}\right) \!,  \label{SigmaV}
\end{equation}%
\begin{equation}
\!\!\frac{2ik_{\parallel }}{a_{1}}\mathcal{E}_{k_{\parallel }k_{\perp
}}^{V}=X_{k_{\parallel }k_{\perp }}^{V}-3\mathcal{E}a_{k_{\parallel
}k_{\perp }}^{V}\!\!+\frac{2}{3}\mu _{k_{\parallel }k_{\perp }}^{V}+3\Sigma 
\overline{\mathcal{H}}_{k_{\parallel }k_{\perp }}^{V}\!\!-\frac{2-k_{\perp
}^{2}}{a_{2}}\mathcal{E}_{k_{\parallel }k_{\perp }}^{T},  \label{epsV}
\end{equation}%
\begin{equation}
\frac{2ik_{\parallel }}{a_{1}}\mathcal{H}_{k_{\parallel }k_{\perp }}^{V}\!\!=%
\frac{1}{a_{2}}\mathcal{H}_{k_{\parallel }k_{\perp }}^{S}\,\!\!-\frac{%
2-k_{\perp }^{2}}{a_{2}}\mathcal{H}_{k_{\parallel }k_{\perp
}}^{T}\,\!\!-3\Sigma \overline{\mathcal{E}}_{k_{\parallel }k_{\perp }}^{V}+3%
\mathcal{E}\overline{\Sigma }_{k_{\parallel }k_{\perp }}^{V},  \label{HV2}
\end{equation}%
\negthinspace 
\begin{equation}
\frac{ik_{\parallel }}{a_{1}}\Sigma _{k_{\parallel }k_{\perp }}^{T}=\frac{%
\Sigma _{k_{\parallel }k_{\perp }}^{V}}{a_{2}}+\frac{3\Sigma }{2}\zeta
_{k_{\parallel }k_{\perp }}^{T}+\overline{\mathcal{H}}_{k_{\parallel
}k_{\perp }}^{T}\ ,  \label{SigmaT}
\end{equation}%
\begin{equation}
\frac{ik_{\parallel }}{a_{1}}\zeta _{k_{\parallel }k_{\perp
}}^{T}\!\!=\!\left( \!\Sigma \!+\!\frac{\Theta }{3}\!\right) \Sigma
_{k_{\parallel }k_{\perp }}^{T}\!\!+\!\frac{1}{a_{2}}a_{k_{\parallel
}k_{\perp }}^{V}\!\!-\!\mathcal{E}_{k_{\parallel }k_{\perp }}^{T},
\label{shearT}
\end{equation}%
\begin{equation}
p_{k_{\parallel }k_{\perp }}^{V}=-\left( \mu +p\right) \mathcal{A}%
_{k_{\parallel }k_{\perp }}^{V}\ ,  \label{pvAv}
\end{equation}%
\begin{equation}
\frac{k_{\perp }^{2}}{a_{2}}\overline{\Sigma }_{k_{\parallel }k_{\perp
}}^{V}=-3\Sigma \xi _{k_{\parallel }k_{\perp }}^{S}+\mathcal{H}%
_{k_{\parallel }k_{\perp }}^{S}\ ,  \label{SigmabarV2}
\end{equation}%
\begin{equation}
\frac{2\!-\!k_{\perp }^{2}}{a_{2}}\!\zeta _{k_{\parallel }k_{\perp
}}^{T}\!\!-\frac{\!\phi _{k_{\parallel }k_{\perp }}^{S}}{a_{2}}\!=\left(
\!\Sigma \!-\!\frac{2\Theta }{3}\!\right) \!\Sigma _{k_{\parallel }k_{\perp
}}^{V}+2\mathcal{E}_{k_{\parallel }k_{\perp }}^{V}\!,  \label{phiS}
\end{equation}%
\begin{equation}
2\overline{\mathcal{H}}_{k_{\parallel }k_{\perp
}}^{V}\!\!=\!\!V_{k_{\parallel }k_{\perp }}^{V}\!\!-\frac{2}{3}%
W_{k_{\parallel }k_{\perp }}^{V}\!+\!\frac{2-k_{\perp }^{2}}{a_{2}}\Sigma
_{k_{\parallel }k_{\perp }}^{T}.  \label{HbarV2}
\end{equation}

We have checked the consistency of the evolution equations and constraints
(Eqs. (\ref{pvAs}), (\ref{AvAs}) and (\ref{pvAv}), containing no evolutions
for $\mathcal{A}_{k_{\parallel }k_{\perp }}^{S}$, $\mathcal{A}_{k_{\parallel
}k_{\perp }}^{V}$ and $p_{k_{\parallel }k_{\perp }}^{V}$) by showing that
the time derivatives of the latter are identities. In fact not all algebraic
relations are independent, their relations are summarised as follows: 1) Eq.
(\ref{pvAs}) follows from Eqs. (\ref{AvAs}) and (\ref{pvAv}); 2) Eq. (\ref%
{SigmabarV1}) follows from Eqs. (\ref{HV}), (\ref{ksziS}), (\ref{HS}) and (%
\ref{SigmabarV2}); 3) Eq. (\ref{phiS}) follows from Eqs. (\ref{HbarV1}), (%
\ref{VW}), (\ref{SigmaT}) and (\ref{HbarV2}); 4) Eq. (\ref{HV2}) follows
from Eqs. (\ref{SigmabarT}), (\ref{ksziS2}), (\ref{HV}) and (\ref{SigmabarV2}%
); 5) Eq. (\ref{ksziS2}) is consequence of Eqs. (\ref{epsbarV1}), (\ref%
{shearbarT1}), (\ref{HV}), (\ref{ksziS}), (\ref{HS}) and (\ref{SigmabarV2});
and finally 6) Eq. (\ref{phiS2}) is consequence of Eqs. (\ref{HbarV1}), (\ref%
{VW}), (\ref{SigmaV}), (\ref{epsV}), (\ref{SigmaT}), (\ref{shearT}) and (\ref%
{HbarV2}). Thus there are 17 constraints for the 28 variables ($\mathcal{A}%
_{k_{\parallel }k_{\perp }}^{S}$, $\mathcal{H}_{k_{\parallel }k_{\perp
}}^{S} $, $\!\phi _{k_{\parallel }k_{\perp }}^{S}$, $\xi _{k_{\parallel
}k_{\perp }}^{S}$, $\mu _{k_{\parallel }k_{\perp }}^{V}$, $p_{k_{\parallel
}k_{\perp }}^{V}$, $\mathcal{A}_{k_{\parallel }k_{\perp }}^{V}$, $%
V_{k_{\parallel }k_{\perp }}^{V}$, $W_{k_{\parallel }k_{\perp }}^{V}$, $%
X_{k_{\parallel }k_{\perp }}^{V}$, $\Sigma _{k_{\parallel }k_{\perp }}^{V}$, 
$\overline{\Sigma }_{k_{\parallel }k_{\perp }}^{V}$, $a_{k_{\parallel
}k_{\perp }}^{V}$, $\overline{a}_{k_{\parallel }k_{\perp }}^{V}$, $\alpha
_{k_{\parallel }k_{\perp }}^{V}$, $\overline{\alpha }_{k_{\parallel
}k_{\perp }}^{V}$, $\mathcal{E}_{k_{\parallel }k_{\perp }}^{V}$, $\overline{%
\mathcal{E}}_{k_{\parallel }k_{\perp }}^{V}$, $\mathcal{H}_{k_{\parallel
}k_{\perp }}^{V} $, $\overline{\mathcal{H}}_{k_{\parallel }k_{\perp }}^{V}$, 
$\Sigma _{k_{\parallel }k_{\perp }}^{T}$, $\overline{\Sigma }_{k_{\parallel
}k_{\perp }}^{T}$, $\zeta _{k_{\parallel }k_{\perp }}^{T}$, $\overline{\zeta 
}_{k_{\parallel }k_{\perp }}^{T}$, $\mathcal{E}_{k_{\parallel }k_{\perp
}}^{T}$, $\overline{\mathcal{E}}_{k_{\parallel }k_{\perp }}^{T}$, $\mathcal{H%
}_{k_{\parallel }k_{\perp }}^{T}$, $\overline{\mathcal{H}}_{k_{\parallel
}k_{\perp }}^{T}$).

Imposing an equation of state $p=p(\mu )$ gives $p_{k_{\parallel }k_{\perp
}}^{V}=c_{s}^{2}\mu _{k_{\parallel }k_{\perp }}^{V}$ where $%
c_{s}^{2}=dp/d\mu $ is the adiabatic speed of sound squared. With the
freedom in the choice of frame (Appendix \ref{Frame}) the condition $a_{a}=0$
can be set. In this case there are 19 constraints for 25 variables. By the
frame choice and Eq. (\ref{ksziS}) we have $\xi _{k_{\parallel }k_{\perp
}}^{S}=0$. The evolution of all other harmonic coefficient follows from the
evolutions of $\mu _{k_{\parallel }k_{\perp }}^{V}$, $\Sigma _{k_{\parallel
}k_{\perp }}^{T}$, $\mathcal{E}_{k_{\parallel }k_{\perp }}^{T}$, $\overline{%
\mathcal{E}}_{k_{\parallel }k_{\perp }}^{T}$, $\mathcal{H}_{k_{\parallel
}k_{\perp }}^{T}$ and $\overline{\mathcal{H}}_{k_{\parallel }k_{\perp }}^{T}$
by (\ref{var7}) and the constraints. The perturbation variables coupled to $%
\mu _{k_{\parallel }k_{\perp }}^{V}$, $\Sigma _{k_{\parallel }k_{\perp
}}^{T} $, $\mathcal{E}_{k_{\parallel }k_{\perp }}^{T}$, $\overline{\mathcal{H%
}}_{k_{\parallel }k_{\perp }}^{T}$ are 
\begin{equation}
{\mathcal{A}}_{k_{\parallel }k_{\perp }}^{V}=\frac{a_{1}}{ik_{\parallel }}%
\frac{{\mathcal{A}}_{k_{\parallel }k_{\perp }}^{S}}{a_{2}}=-\frac{c_{s}^{2}}{%
\mu +p}\mu _{k_{\parallel }k_{\perp }}^{V}~,  \label{constrE}
\end{equation}%
\begin{equation}
\frac{ik_{\parallel }}{a_{1}}\zeta _{k_{\parallel }k_{\perp }}^{T}=\left(
\Sigma +\frac{\Theta }{3}\right) \Sigma _{k_{\parallel }k_{\perp }}^{T}-{%
\mathcal{E}}_{k_{\parallel }k_{\perp }}^{T}~,
\end{equation}%
\begin{equation}
\frac{ik_{\parallel }}{a_{1}a_{2}}\Sigma _{k_{\parallel }k_{\perp
}}^{V}=-\left( B+\frac{2-k_{\perp }^{2}}{a_{2}^{2}}\right) \frac{\Sigma
_{k_{\parallel }k_{\perp }}^{T}}{2}+\frac{3}{2}\Sigma {\mathcal{E}}%
_{k_{\parallel }k_{\perp }}^{T}-i\frac{k_{\parallel }}{a_{1}}\overline{%
\mathcal{H}}_{k_{\parallel }k_{\perp }}^{T}~,
\end{equation}%
\begin{equation}
\frac{2\overline{\mathcal{H}}_{k_{\parallel }k_{\perp }}^{V}}{3a_{2}}=-\frac{%
\Sigma }{a_{2}B}\mu _{k_{\parallel }k_{\perp }}^{V}+\Sigma C{\mathcal{E}}%
_{k_{\parallel }k_{\perp }}^{T}-{\mathcal{E}}\Sigma _{k_{\parallel }k_{\perp
}}^{T}-\frac{ik_{\parallel }}{a_{1}}J\frac{\overline{\mathcal{H}}%
_{k_{\parallel }k_{\perp }}^{T}}{3}~,  \label{HbarVconstr}
\end{equation}%
\begin{equation}
\frac{V_{k_{\parallel }k_{\perp }}^{V}}{a_{2}}=-\frac{2\Sigma }{a_{2}B}\mu
_{k_{\parallel }k_{\perp }}^{V}-\frac{2B}{3}\left( 2+C\right) \Sigma
_{k_{\parallel }k_{\perp }}^{T}+\Sigma \left( 1+2C\right) \mathcal{E}%
_{k_{\parallel }k_{\perp }}^{T}-\frac{2}{3}\frac{ik_{\parallel }}{a_{1}}%
\left( 1+J\right) \overline{\mathcal{H}}_{k_{\parallel }k_{\perp }}^{T}~,
\label{VVconstr}
\end{equation}%
\begin{eqnarray}
\frac{W_{k_{\parallel }k_{\perp }}^{V}}{a_{2}} &=&\frac{3\Sigma }{2a_{2}B}%
\mu _{k_{\parallel }k_{\perp }}^{V}+\left( 3{\mathcal{E}}+\frac{2-k_{\perp
}^{2}}{a_{2}^{2}}-B\right) \frac{\Sigma _{k_{\parallel }k_{\perp }}^{T}}{2} 
\nonumber \\
&&\!\!\!\!\!\!\!\!\!\!+\frac{3\Sigma }{2}\left( 1-C\right) \mathcal{E}%
_{k_{\parallel }k_{\perp }}^{T}-\frac{ik_{\parallel }}{2a_{1}}\left(
2-J\right) \overline{\mathcal{H}}_{k_{\parallel }k_{\perp }}^{T}~,
\label{WVconstr}
\end{eqnarray}%
\begin{equation}
\frac{ik_{\parallel }}{a_{1}}\frac{\phi _{k_{\parallel }k_{\perp }}^{S}}{%
a_{2}^{2}}=-\frac{k_{\parallel }^{2}}{a_{1}^{2}}\frac{2}{a_{2}B}\mu
_{k_{\parallel }k_{\perp }}^{V}-\frac{BL}{3\Sigma }\Sigma _{k_{\parallel
}k_{\perp }}^{T}+\left( L-\frac{2-k_{\perp }^{2}}{a_{2}^{2}B}\frac{k_{\perp
}^{2}}{a_{2}^{2}}\right) \mathcal{E}_{k_{\parallel }k_{\perp }}^{T}-\frac{%
ik_{\parallel }}{a_{1}}\frac{2L}{3\Sigma }\overline{\mathcal{H}}%
_{k_{\parallel }k_{\perp }}^{T}~,
\end{equation}%
\begin{eqnarray}
\frac{ik_{\parallel }}{a_{1}a_{2}}\alpha _{k_{\parallel }k_{\perp }}^{V} &=&-%
\left[ \frac{3\Sigma }{2B}+\left( \Sigma +\frac{\Theta }{3}\right) \frac{%
c_{s}^{2}}{\mu +p}\right] \frac{\mu _{k_{\parallel }k_{\perp }}^{V}}{a_{2}} 
\nonumber \\
&&-\frac{3\mathcal{E}}{2}\Sigma _{k_{\parallel }k_{\perp }}^{T}+\frac{%
3\Sigma }{2}C\mathcal{E}_{k_{\parallel }k_{\perp }}^{T}-\frac{ik_{\parallel }%
}{2a_{1}}J\overline{\mathcal{H}}_{k_{\parallel }k_{\perp }}^{T}~,
\end{eqnarray}%
\begin{eqnarray}
\frac{X_{k_{\parallel }k_{\perp }}^{V}}{a_{2}} &=&\left[ 1-\frac{3}{B}\left( 
\frac{k_{\perp }^{2}}{a_{2}^{2}}+3\mathcal{E}\right) \right] \frac{\mu
_{k_{\parallel }k_{\perp }}^{V}}{3a_{2}}-\frac{\mathcal{E}}{\Sigma }\left( 
\frac{2}{a_{2}^{2}}+3\mathcal{E}\right) \Sigma _{k_{\parallel }k_{\perp
}}^{T}+C\left( \frac{k_{\perp }^{2}}{a_{2}^{2}}+3\mathcal{E}\right) \mathcal{%
E}_{k_{\parallel }k_{\perp }}^{T}  \nonumber \\
&&+\frac{a_{1}}{ik_{\parallel }}\left[ \mathcal{E}\left( \mathcal{E}+\frac{2%
}{3a_{2}^{2}}\right) \frac{k_{\parallel }^{2}}{a_{1}^{2}}-\Sigma ^{2}\frac{%
2-k_{\perp }^{2}}{4a_{2}^{2}}\frac{k_{\perp }^{2}}{a_{2}^{2}}\right] \frac{6%
\overline{\mathcal{H}}_{k_{\parallel }k_{\perp }}^{T}}{\Sigma B}~,
\label{XVconstr}
\end{eqnarray}%
\begin{eqnarray}
\frac{ik_{\parallel }}{a_{1}a_{2}}\mathcal{E}_{k_{\parallel }k_{\perp
}}^{V}\! &=&\frac{k_{\parallel }^{2}}{a_{1}^{2}}\frac{\mu _{k_{\parallel
}k_{\perp }}^{V}}{a_{2}B}\!\!-\frac{3\mathcal{E}}{2}\left( \Sigma +\frac{%
\Theta }{3}\right) \Sigma _{k_{\parallel }k_{\perp }}^{T}+\left( \frac{3%
\mathcal{E}}{2}-C\frac{k_{\parallel }^{2}}{a_{1}^{2}}\right) \mathcal{E}%
_{k_{\parallel }k_{\perp }}^{T}  \nonumber \\
&&-\frac{ik_{\parallel }}{a_{1}}\frac{3\overline{\mathcal{H}}_{k_{\parallel
}k_{\perp }}^{T}}{2B}\left[ 2\mathcal{E}\left( \Sigma +\frac{\Theta }{3}%
\right) +\Sigma \frac{2-k_{\perp }^{2}}{a_{2}^{2}}\right] ~,
\end{eqnarray}%
and which are coupled to $\overline{\mathcal{E}}_{k_{\parallel }k_{\perp
}}^{T}$, $\mathcal{H}_{k_{\parallel }k_{\perp }}^{T}$ are%
\begin{eqnarray}
{\mathcal{H}}_{k_{\parallel }k_{\perp }}^{V} &=&\frac{ik_{\parallel }}{a_{1}}%
\frac{a_{2}}{k_{\perp }^{2}}{\mathcal{H}}_{k_{\parallel }k_{\perp }}^{S}=-%
\frac{ik_{\parallel }}{a_{1}}\overline{\alpha }_{k_{\parallel }k_{\perp
}}^{V}=\frac{ik_{\parallel }}{a_{1}}\overline{\Sigma }_{k_{\parallel
}k_{\perp }}^{V}  \nonumber \\
&=&\frac{2-k_{\perp }^{2}}{2a_{2}}\overline{\Sigma }_{k_{\parallel }k_{\perp
}}^{T}=\frac{2-k_{\perp }^{2}}{a_{2}B}\left( \frac{3\Sigma }{2}\overline{%
\mathcal{E}}_{k_{\parallel }k_{\perp }}^{T}+\frac{ik_{\parallel }}{a_{1}}{%
\mathcal{H}}_{k_{\parallel }k_{\perp }}^{T}\right) ~,
\end{eqnarray}%
\begin{equation}
\frac{B}{2}\overline{\zeta }_{k_{\parallel }k_{\perp }}^{T}=\left( \Sigma +%
\frac{\Theta }{3}\right) {\mathcal{H}}_{k_{\parallel }k_{\perp }}^{T}-\frac{%
a_{1}}{ik_{\parallel }}\left( \frac{k_{\parallel }^{2}}{a_{1}^{2}}-\frac{%
2-k_{\perp }^{2}}{2a_{2}^{2}}\right) \overline{\mathcal{E}}_{k_{\parallel
}k_{\perp }}^{T}~,
\end{equation}%
\begin{equation}
\overline{\mathcal{E}}_{k_{\parallel }k_{\perp }}^{V}=\frac{2-k_{\perp }^{2}%
}{2a_{2}}\left[ \frac{a_{1}}{ik_{\parallel }}\left( 1-\frac{9\Sigma ^{2}}{2B}%
\right) \overline{\mathcal{E}}_{k_{\parallel }k_{\perp }}^{T}-\frac{3\Sigma 
}{B}\mathcal{H}_{k_{\parallel }k_{\perp }}^{T}\right] ~,  \label{constrF}
\end{equation}

where $B$ and $C$ are defined in Eqs. (\ref{Bdef}) and (\ref{Cdef}),
respectively, and%
\begin{equation}
L\equiv 2C\frac{k_{\parallel }^{2}}{a_{1}^{2}}+\frac{k_{\perp }^{2}}{%
a_{2}^{2}}\left( 1+\frac{2-k_{\perp }^{2}}{a_{2}^{2}B}\right) ~,
\end{equation}%
\begin{equation}
J\equiv \frac{(2-k_{\perp }^{2})k_{\perp }^{2}a_{1}^{2}}{k_{\parallel
}^{2}a_{2}^{4}B}+2C~.
\end{equation}


\begin{thebibliography}{99}
\bibitem{Komatsu49} E. Komatsu et.al., \textit{First-Year Wilkinson
Microwave Anisotropy Probe (WMAP) Observations: Tests of Gaussianity}, 
\textit{Astrophys. J. Suppl.} \textbf{148} (2003) 119
[arXiv:astro-ph/0302223].

\bibitem{Spergel48} D.N. Spergel et. al., \textit{Three-Year Wilkinson
Microwave Anisotropy Probe (WMAP) Observations: Implications for Cosmology}, 
\textit{Astrophys. J.} \textbf{170} (2007) 377 [astro-ph/0603449].

\bibitem{WMAP9yr} G. Hinshaw et al., \textit{Nine-year Wilkinson Microwave
Anisotropy Probe (WMAP) Observations: Cosmological Parameter Results}, 
\textit{Astrophys. J. Suppl.} \textbf{208} (2013) 19 [arXiv:1212.5226].

\bibitem{Planck1} Planck collaboration, P.A.R. Ade et al., \textit{Planck
2013 results. XVI. Cosmological parameters}, \textit{Astron. Astrophys. }%
\textbf{571} (2014) A16 [arXiv:1303.5076].

\bibitem{Planck2} Planck collaboration, P.A.R. Ade et al., \textit{Planck
2015 results. XIII. Cosmological parameters}, \textit{Astron. Astrophys. }%
(2015) [arXiv:1502.01589].

\bibitem{Bennett46} C.L. Bennett et. al., \textit{First-Year Wilkinson
Microwave Anisotropy Probe (WMAP) Observations: Preliminary Maps and Basic
Results}, \textit{Astrophys. J. Suppl.} \textbf{148} (2003) 1
[astro-ph/0302207].

\bibitem{Oliveria45} A. de Oliveira-Costa, M. Tegmark, M. Zaldarriaga and A.
Hamilton, \textit{Significance of the largest scale CMB fluctuations in WMAP}%
, \textit{Phys. Rev. D} \textbf{69} (2004) 063516 [astro-ph/0307282].

\bibitem{Vielva59} P. Vielva et. al., \textit{Detection of Non-Gaussianity
in the Wilkinson Microwave Anisotropy Probe First-Year Data Using Spherical
Wavelets}, \textit{Astrophys. J.} \textbf{609} (2004) 22 [astro-ph/0310273].

\bibitem{PlanckAnomaly} Planck collaboration, P.A.R. Ade et al., \textit{%
Planck 2015 results. XVI. Isotropy and statistics of the CMB}, \textit{%
Astron. Astrophys. }(2015) [arXiv:1506.07135].

\bibitem{alt1} T. Padmanabhan, \textit{Accelerated expansion of the universe
driven by tachyonic matter}, \textit{Phys. Rev. D} \textbf{66} (2002) 021301
[hep-th/0204150].

\bibitem{alt2} A.V. Frolov, L. Kofman and A.A. Starobinsky, \textit{%
Prospects and problems of tachyon matter cosmology}, \textit{Phys. Lett. B} 
\textbf{545} (2002) 8 [hep-th/0204187].

\bibitem{alt3} C. Cs\'{a}ki, N. Kaloper and J. Terning, \textit{The
accelerated acceleration of the Universe}, JCAP \textbf{06} (2006) 022
[astro-ph/0507148].

\bibitem{alt4} V. Gorini, A.Y. Kamenshchik, U. Moschella and V. Pasquier, 
\textit{Tachyons, scalar fields and cosmology}, \textit{Phys. Rev. D} 
\textbf{69} (2004) 123512 [hep-th/0311111].

\bibitem{alt5} C.-M. Yoo, K.-I. Nakao, M. Sasaki, \textit{CMB observations
in LTB universes: Part I: Matching peak positions in the CMB spectrum}, 
\textit{JCAP} \textbf{07} (2010) 012 [arXiv:1005.0048].

\bibitem{alt6} N. Shin'Ichi and S.D. Odintsov, \textit{Unified cosmic
history in modified gravity: From F(R) theory to Lorentz non-invariant models%
}, \textit{Phys. Rep.} \textbf{505} (2011) 59 [arXiv:1011.0544].

\bibitem{alt7} L. Lombriser, A. Slosar, U. Seljak and W. Hu, \textit{%
Constraints on f(R) gravity from probing the large-scale structure}, \textit{%
Phys. Rev. D} \textbf{85} (2012) 124038 [arXiv:1003.3009].

\bibitem{alt8} S. Tsujikawa, \textit{Quintessence: A Review}, \textit{Class.
Quant. Grav.} \textbf{30} (2013) 214003 [arXiv:1304.1961].

\bibitem{alt9} B. Novosyadlyj, O. Sergijenko, R. Durrer and V. Pelykh, 
\textit{Constraining the dynamical dark energy parameters: Planck-2013 vs
WMAP9}, JCAP \textbf{05} (2014) 030 [arXiv:1312.6579].

\bibitem{alt10} Z. Keresztes and L. \'{A}. Gergely, Combined cosmological
tests of a bivalent tachyonic dark energy scalar field model, JCAP \textbf{11%
} (2014) 026 [arXiv:1408.3736].

\bibitem{alt11} L. \'{A}. Gergely and S. Tsujikawa, \textit{Effective field
theory of modified gravity with two scalar fields: dark energy and dark
matter}, \textit{Phys. Rev. D} \textbf{89} (2014) 064059 [arXiv:1402.0553].

\bibitem{alt12} T. Denkiewicz, \textit{Dark energy and dark matter
perturbations in singular universes}, \textit{JCAP} \textbf{03} (2015) 037
[arXiv:1411.6169].

\bibitem{H1} M.L. McClure and C.C. Dyer, \textit{Anisotropy in the Hubble
constant as observed in the HST extragalactic distance scale key project
results}, \textit{New Astronomy} \textbf{12} (2007) 533 [astro-ph/0703556].

\bibitem{H2} D.L. Wiltshire, P.R. Smale, T. Mattsson and R. Watkins, \textit{%
Hubble flow variance and the cosmic rest frame}, \textit{Phys. Rev. D} 
\textbf{88} (2013) 083529 [arXiv:1201.5371].

\bibitem{Dec} R.-G. Cai and Z.-L. Tuo, \textit{Direction dependence of the
deceleration parameter}, \textit{JCAP} \textbf{02} (2012) 004
[arXiv:1109.0941].

\bibitem{Doroschkevich} A.G. Doroshkevich, Ya. B. Zel'dovich and I.D.
Novikov, \textit{Perturbations in an anisotropic homogeneous universe}, 
\textit{Zh. Ehksp. Teor. Fiz.} \textbf{60} (1971) 3 [Sov. Phys. \textit{JETP} \textbf{33} (1971) 1].

\bibitem{Perko} T.E. Perko, A. Matzner and L.C. Shepley, \textit{Galaxy
Formation in Anisotropic Cosmologies}, \textit{Phys. Rev. D} \textbf{6}
(1972) 969.

\bibitem{Hu} B.L. Hu and T. Regge, \textit{Perturbations on the Mixmaster
Universe}, \textit{Phys. Rev. Lett.} \textbf{29} (1972) 1616.

\bibitem{Abbott} R.B. Abbott, B. Bednarz and S.D. Ellis, \textit{%
Cosmological perturbations in Kaluza-Klein models}, \textit{Phys Rev. D} 
\textbf{33} (1986) 2147.

\bibitem{Tomita} K. Tomita and M. Den, \textit{Gauge-invariant perturbations
in anisotropic homogeneous cosmological models}, \textit{Phys. Rev. D} 
\textbf{34} (1986) 3570.

{\color{black}

\bibitem{Gumruk} A.E. G\"{u}mr\"{u}k\c{c}\"{u}o\u{g}lu, C.R. Contaldi and M.
Peloso, \textit{\ Inflationary perturbations in anisotropic background and
their imprint on the cosmic microwave background}, \textit{JCAP} \textbf{11}
(2007) 005 [astro-ph/0707.4179]

\bibitem{Pereira} T.S. Periera, C. Pitrou and J.-P. Uzan, \textit{\ Theory
of cosmological perturbations in an anisotropic universe} \textit{JCAP} 
\textbf{09} (2007) 006 [astro-ph/0707.0736] 

\bibitem{Pitrou} C. Pitrou, T.S. Periera and J.-P. Uzan, \textit{\
Predictions from an anisotropic inflationary era} \textit{JCAP} \textbf{04}
(2008) 004 [astro-ph/0801.3596] 

}

\bibitem{Lifshitz} E.M. Lifshitz and I.M. Khalatnikov, \textit{%
Investigations in relativistic cosmology}, \textit{Adv. Phys.} \textbf{12}
(1963) 185.

\bibitem{Bardeen} J.M. Bardeen, \textit{Gauge-invariant cosmological
perturbations}, \textit{Phys. Rev. D} \textbf{22} (1980) 1882.

\bibitem{Stewart} J.M. Stewart, \textit{Perturbations of
Friedmann-Robertson-Walker cosmological models}, \textit{Class. Quantum Grav.%
} \textbf{7} (1990) 1169.

\bibitem{Hawking} S.W. Hawking, \textit{Perturbations of an Expanding
Universe}, \textit{Astrophys. J.} \textbf{145} (1966) 544.

\bibitem{Olson} D.W. Olson, \textit{Density perturbations in cosmological
models}, \textit{Phys. Rev. D} \textbf{14} (1976) 327.

\bibitem{cov1} G.F.R Ellis and M. Bruni, \textit{Covariant and
gauge-invariant approach to cosmological density fluctuations}, \textit{%
Phys. Rev. D} \textbf{40} (1989) 1804.

\bibitem{cov2} P.K.S.~Dunsby, \textit{Gauge invariant perturbations in
multi-component fluid cosmologies}, \textit{Class.\ Quant.\ Grav.}\ \textbf{8%
} (1991) 1785.

\bibitem{cov3} M.~Bruni, P.K.S.~Dunsby and G.F.R.~Ellis, \textit{%
Cosmological perturbations and the physical meaning of gauge-invariant
variables}, \textit{Astrophys.\ J.}\ \textbf{395} (1992) 34.

\bibitem{cov4} P.K.S.~Dunsby, M.~Bruni and G.F.R.~Ellis, \textit{Covariant
perturbations in a multifluid cosmological medium}, \textit{Astrophys.\ J.}\ 
\textbf{395} (1992) 54.

\bibitem{cov5} P.K.S.~Dunsby, \textit{Covariant perturbations of anisotropic
cosmological models}, \textit{Phys.\ Rev.\ D} \textbf{48} (1993) 3562.

\bibitem{cov6} B. Osano, Beyond the standard model of cosmology: a perturbative approach, PhD Thesis (University of Cape Town, 2008).

\bibitem{cov7} B. Osano, \textit{The Decoupling of Scalar-Modes from a
Linearly Perturbed Dust-Filled Bianchi Type-I Model}, \textit{Chin. Phys.
Lett.} \textbf{31} (2014) 010402 [arXiv:1504.01900].

{\color{black}
\bibitem{cov-applications1} P.~K.~S.~Dunsby, \textit{A fully covariant description of CMB anisotropies}, Class.\ Quant.\ Grav.\ \textbf{14}, 3391 (1997) [gr-qc/9707022].

\bibitem{cov-applications2} T.~Gebbie, P.~Dunsby and G.~F.~R.~Ellis, \textit{1+3 Covariant Cosmic Microwave Background anisotropies II: 
The almost - Friedmann Lemaitre model}, Annals Phys.\ \textbf{282}, 321 (2000) [astro-ph/9904408].

\bibitem{cov-applications3} M.~Marklund, P.~K.~S.~Dunsby, M.~Servin, G.~Betschart and C.~Tsagas, 
\textit{Charged multifluids in general relativity}, Class.\ Quant.\ Grav.\ \textbf{20}, 1823 (2003) [gr-qc/0211067]. 

\bibitem{cov-applications4} C.~G.~Tsagas, P.~K.~S.~Dunsby and M.~Marklund, 
\textit{Gravitational wave amplification of seed magnetic fields}, Phys.\ Lett.\ B \textbf{561}, 17 (2003) [astro-ph/0112560].

\bibitem{cov-applications5} A.~Abebe, M.~Abdelwahab, A.~de la Cruz-Dombriz and P.~K.~S.~Dunsby, 
\textit{Covariant gauge-invariant perturbations in multifluid f(R) gravity}, Class.\ Quant.\ Grav.\ \textbf{29}, 135011 (2012) [arXiv:1110.1191].
}

\bibitem{StewartWalker} J.M. Stewart and M. Walker, \textit{Perturbations of
space-times in general relativity}, \textit{Proc. R. Soc. London} \textbf{%
A341} (1974) 49.

\bibitem{KASAscalar} M. Bradley, P.K.S. Dunsby, M. Forsberg and Z.
Keresztes, \textit{Density growth in Kantowski-Sachs cosmologies with
cosmological constant}, \textit{Class. Quantum Grav. }\textbf{29} (2012)
095023 [arXiv:1106.4932].

\bibitem{Cargese} G.F.R. Ellis and H. van Elst: \textit{Cosmological models
in Theoretical and Observational Cosmology}, in \textit{NATO Adv. Study
Inst. Ser. C. Math. Phys. Sci.} \textbf{541} (1999) 1, ed. M. Lachi\'{e}%
ze-Rey, Kluwer Acad., Dordrecht [gr-qc/9812046].

\bibitem{perturb2} G.F.R. Ellis, M. Bruni and J. Hwang, \textit{Covariant
and gauge-independent perfect-fluid Robertson-Walker perturbations}, \textit{%
Phys. Rev. D} \textbf{40} (1989) 1819.

\bibitem{perturb4} A. Challinor and A. Lasenby, \textit{Cosmic Microwave
Background Anisotropies in the Cold Dark Matter Model: A Covariant and
Gauge-invariant Approach}, Astrophys. J. \textbf{513} (1999) 1
[astro-ph/9804301].

\bibitem{LRS} H. van Elst and G.F.R. Ellis, \textit{The Covariant Approach
to LRS Perfect Fluid Spacetime Geometries}, \textit{Class. Quantum Grav.} 
\textbf{13} (1996) 1099 [gr-qc/9510044].

\bibitem{Schperturb} C.A. Clarkson and R.K. Barrett, \textit{Covariant
Perturbations of Schwarzschild Black Holes},\textit{\ Class. Quantum Grav.} 
\textbf{20} (2003) 3855 [gr-qc/0209051].

\bibitem{Schperturb2} C.~A.~Clarkson, M.~Marklund, G.~Betschart and P.~K.~S.~Dunsby, Astrophys.\ J.\ \textbf{613}, 492 (2004).

\bibitem{LRSIIscalar} G. Betschart and C. Clarkson, \textit{Scalar field and
electromagnetic perturbations on Locally Rotationally Symmetric spacetimes}, 
\textit{Class. Quantum Grav.} \textbf{21} (2004) 5587 [gr-qc/0404116].

\bibitem{LRSIItensor} R.B. Burston, \textit{1+1+2 gravitational
perturbations on LRS class II space-times: Decoupling GEM tensor harmonic
amplitudes}, \textit{Class. Quantum Grav.} \textbf{25} (2008) 075004
[arXiv:0708.1812].

\bibitem{1+1+2} C. Clarkson, \textit{A covariant approach for perturbations
of rotationally symmetric spacetimes}, \textit{Phys. Rev.\ D} \textbf{76}
(2007) 104034 [arXiv:0708.1398].

\bibitem{Bonometto} G.F.R. Ellis, \textit{Cosmological Models}, in \textit{\
Modern Cosmology}, ed. S. Bonometto, V. Gorini and U. Moschella, IOP
Publishing Ltd (2002).

\bibitem{Tsagas} C.G. Tsagas, A. Challinor and R. Maartens, \textit{\
Relativistic cosmology and large-scale structure}, \textit{Phys. Rept.} 
\textbf{465} (2008) 61 [arXiv:0705.4397].

\textcolor{black}{
\bibitem{HawkingEllis} S.W. Hawking, G.F.R. Ellis, \textit{
The large scale structure of space-time}, Cambridge Univ. Press. 1973.}

\bibitem{3+1+1} Z. Keresztes and L. \'{A}. Gergely, \textit{Covariant
gravitational dynamics in 3+1+1 dimensions}, \textit{Class. Quantum Grav.} 
\textbf{27} (2010) 105009 [arXiv:0909.0490].

\bibitem{MarklundBradley} M. Marklund and M. Bradley, \textit{Invariant
construction of solutions to Einstein's field equations - LRS perfect fluids
II}, \textit{Class. Quantum Grav.} \textbf{16} (1999) 1577 [gr-qc/9808062].

\bibitem{Brunietal} M. Bruni, S. Matarrese, S. Mollerach and S. Sonego, 
\textit{Perturbations of spacetime: gauge transformations and gauge
invariance at second order and beyond}, \textit{Class. Quantum Grav.} 
\textbf{14} (1997) 2585 [gr-qc/9609040].

\bibitem{BruniSonego} M. Bruni and S. Sonego, \textit{Observables and gauge
invariance in the theory of non-linear spacetime perturbations}, \textit{
Class. Quantum Grav.} \textbf{16} (1999) L29 [gr-qc/9906017].

\textcolor{black}{
\bibitem{Regge} T.~Regge and J.~A.~Wheeler, \textit{Stability of a
Schwarzschild Singularity}, \textit{Phys.\ Rev.}\ \textbf{108}, 1063 (1957).
}

\textcolor{black}{
\bibitem{Zerilli} F. J. Zerilli, \textit{Effective Potential for Even-Parity
Regge-Wheeler Gravitational Perturbation Equations}, \textit{Phys. Rev.
Lett. }\textbf{24}, 737 (1970).
}

\textcolor{black}{
\bibitem{Zerilli2} F.~J. Zerilli, \textit{Perturbation analysis for
gravitational and electromagnetic radiation in a Reissner-Nordstr\"{o}m
geometry}, \textit{Phys.\ Rev.}\ D \textbf{9}, 860 (1974).
}

\bibitem{Isaacson1} R.A. Isaacson, \textit{Gravitational Radiation in the
Limit of High Frequency. I. The Linear Approximation and Geometrical Optics}
, \textit{Phys. Rev.} \textbf{166} (1968) 1263.

\bibitem{Isaacson2} R.A. Isaacson, \textit{Gravitational Radiation in the
Limit of High Frequency. II. Nonlinear Terms and the Effective Stress Tensor}
, \textit{Phys. Rev.} \textbf{166} (1968) 1272.

\textcolor{black}{
\bibitem{Thorne} K.P. Thorne,
\textit{Multipole expansions of gravitational radiation},
\textit{Rev. Mod. Phys.} \textbf{52} (1980) 299.}

\textcolor{black}{
\bibitem{Harrison} E.R. Harrison, \textit{Normal modes of vibrations of the universe},
\textit{Rev. Mod. Phys.}\textbf{39} (1967) 862.}
\end{thebibliography}
\end{document}